\documentclass[%
reprint,
pra,
superscriptaddress
]{revtex4-1}

\usepackage{physics}
\usepackage{ascmac}
\usepackage{graphicx}% Include figure files
\usepackage{dcolumn}% Align table columns on decimal point
\usepackage{bm}% bold math
\usepackage{braket}
\usepackage{here}
\usepackage{wrapfig}
\usepackage[dvipsnames]{xcolor}

\allowdisplaybreaks

\begin{document}

%\preprint{APS/123-QED}

\title{
Collective excitations of a Bose-condensed gas: Fate of second sound in the crossover regime between hydrodynamic and collisionless regimes
}% Force line breaks with \\

\author{Hoshu Hiyane}%
\email{hoshu.hiyane@oist.jp}
\affiliation{Quantum Systems Unit, Okinawa Institute of Science and Technology Graduate University, Onna, Okinawa 904-0495, Japan\\
}

\author{Shohei Watabe}%
\affiliation{%
  Department of Physics, Tokyo University of Science, Shinjuku, Tokyo, 162-8601, Japan\\
}%
\affiliation{%
   Faculty of Engineering, Computer Science and Engineering, Shibaura Institute of Technology, Toyosu, Tokyo, 135-8548, Japan\\
}%

\author{Tetsuro Nikuni}%
\affiliation{%
  Department of Physics, Tokyo University of Science, Shinjuku, Tokyo, 162-8601, Japan\\
}%

\begin{abstract}
We develop the moment method for Bose--Einstein condensates at finite temperatures that enable us to study collective sound modes from the hydrodynamic to the collisionless regime. 
In particular, we investigate collective excitations in a weakly interacting dilute Bose gas by applying the moment method to the Zaremba--Nikuni--Griffin equation, which is the coupled equation of the Boltzmann equation with the generalized Gross--Pitaevskii equation. 
Utilizing the moment method, collective excitations in the crossover regime between the hydrodynamic and collisionless regimes are investigated in detail. 
In the crossover regime, the second sound mode loses the weight of the density response function because of the significant coupling with incoherent modes, whereas the first sound shows a distinct but broad peak structure. 
We compare the result obtained by the moment method with that of the Landau two-fluid equations and show that the collective mode predicted by the Landau two-fluid equations well coincides with the result from the moment method even far from the hydrodynamic regime, whereas clear distinction also emerges in the relatively higher momentum regime.

\end{abstract} 
\maketitle

%\tableofcontents

% -------------------------------------------[ INTRODUCTION ]---------------------------------------------------- %
\section{\label{sec:introduction}INTRODUCTION}
Collective excitation is one of the most fundamental concepts in many-body physics~\cite{stoof2008ultracold,pitaevskii2016bose}. 
For the homogeneous Bose-condensed gas at zero temperature, the Gross--Pitaevskii (GP) equation predicts the Bogoliubov excitation~\cite{pitaevskii2016bose,andrews_97,mappelink}, whose dispersion consists of two main regions: the long-wavelength phonon excitation and the short-wavelength particle-like excitation.
At finite temperatures, the situation is more complicated due to the coexistence of the condensate and noncondensate components. 
In the hydrodynamic region, the two-fluid theory predicts the first and second sounds, where in-phase and out-of-phase modes emerge between condensates and noncondensates. 
The zero-temperature GP equation explains a mean-field type of collisionless mode. 
In a dilute Bose gas, the collisionless regime, where the  Landau two-fluid model is out of scope, is easier to address experimentally.
On the other hand, the liquid $^4$He is inherently incompressible, resulting in the second sound as an entropy wave with the normal fluid and superfluid being in out-of-phase oscillation without involving density fluctuation~\cite{Pitaevskii_Stringari_2017}.

There have been numerous approaches to extend the zero-temperature GP equation or Bogoliubov theory to the finite-temperature Bose--Einstein condensates (BECs)~\cite{davis_01,rooney_12,griffin_96,giorgini_99,giorgini_98,Proukakis_finiteTmodeling}.
In this paper, we study the collective excitation based on the Zaremba--Nikuni--Griffin (ZNG) formalism, which provides coupled equations of motion for finite-temperature Bose-condensed gases~\cite{ZNG_original} extended from the pioneering work by Kirkpatrick and Dorfmann \cite{kirkpatrick_dorfman}. 
A striking feature of this model is its applicability; since the thermal cloud is governed by the Boltzmann equation, this approach is not restricted to either collisionless or hydrodynamic regimes. 

In the hydrodynamic regime, the ZNG formalism provides a microscopic derivation of the Landau--Khalatnikov two-fluid equations including the transport coefficients~\cite{ZNG_original,chapmannenskog}. 
On the other hand, in the collisionless regime, Williams and Griffin applied the static-thermal-cloud approximation to the ZNG equation and derived the finite-temperature Stringari equation~\cite{williamsfiniteTstringari}. 
This equation describes the Bogoliubov excitation with collisional damping~\cite{yellowbook}. 

However, the crossover from the hydrodynamic to the collisionless regime has not yet been studied within a single framework. 
In the hydrodynamic limit, the Boltzmann equation can be reduced to the hydrodynamic equation with a few coarse-grained (or thermodynamic) variables. 
 On the other hand, in the collisionless regime far from the local equilibrium, one requires full knowledge of the distribution function of the gas. This means that one needs a set of equations that contains not a few moments. 
The moment method shows great applicability in addressing the crossover from the hydrodynamic to collisionless regimes. 
This method was introduced to solve the linearized Boltzmann equation exactly~\cite{khalatnikov_abrikosov,brooker_sykes}, and recently refurbished to investigate the crossover from the hydrodynamic to the collisionless regime in the normal systems~\cite{watabeosawanikuni,watabenikuni,asano,narushima}.

In the present paper, we develop the moment method for the ZNG coupled equations and investigate collective modes of finite-temperature Bose--Einstein condensed gases through the density response function. 
By extending the moment method previously developed for the normal systems, we rewrite the linearized ZNG equation in terms of the moments with the relaxation-time approximation. 
We also show that the Landau two-fluid equations can be derived by truncating the hierarchy of the moment equations.
By solving the moment equations numerically, we study the crossover between hydrodynamic and collisionless regimes.
The result from the moment method is compared with the dissipative Landau two-fluid theory and shows that the second sound at the crossover regime significantly couples to the other incoherent modes and loses its weight.
The first sound, on the other hand, smoothly crossovers to the collisionless regime and eventually loses its weight in the response function.

This paper is organized as follows.
In Sec.~\ref{sect:linearzation}, we develop the moment method for the finite temperature BECs.
Extending the moment method previously developed for the normal systems, we rewrite the linearized ZNG equation in terms of the moments.
The relaxation-time approximation in the framework of the moment method will be reviewed in detail.
Section~\ref{sect:two-fluid} is devoted  to the derivation of the Landau two-fluid equation by truncating the hierarchy of the moment equations.
In Sec.~\ref{sect:result}, we solve the moment equations numerically and discuss the crossover between hydrodynamic and collisionless regimes.
In Sec.~\ref{sect:conclusion}, the conclusion of this paper is summarized.

% -------------------------------------------[ Method ]---------------------------------------------------- %
\section{Linearized Zaremba--Nikuni--Griffin equation and moment method}\label{sect:linearzation}

In the framework of the ZNG formalism, the dynamics of the Bose-condensate order parameter $\Psi$ is described by the generalized Gross--Pitaevskii equation (GGPE) \cite{yellowbook}:
\begin{align}
i\hbar\pdv{\Psi(\bm r,t)}{t}
=&\biggl[-\frac{\hbar^2\nabla^2}{2m}+U_{\rm ext}(\bm r,t)+gn_c(\bm r,t)\notag\\
&+2g\tilde {n}(\bm r,t)-iR(\bm r,t)
\biggr]
\Psi(\bm r,t).
\end{align}
The dynamics of the noncondensate atoms is described by the semiclassical distribution function $f(\bm r,\bm p,t)$, which obeys the Boltzmann equation~\cite{kirkpatrick_dorfman,ZNG_original}:
\begin{align}
	&\bqty{
 \frac{\partial }{\partial t}
+\frac{\bm p}{m}\cdot\nabla_{\bm r}
-\nabla_{\bm r} U(\bm r,t)\cdot\nabla_{\bm p}
}
f(\bm p,\bm r,t)\notag\\
&=C_{12}[f,\Psi]+C_{22}[f],
\end{align}
where $n_c(\bm r,t)=|\Psi(\bm r,t)|^2$ is the number density of condensate, and
\begin{align}
	\tilde n(\bm r,t)=\int\frac{d\bm p}{(2\pi\hbar)^3}f(\bm p,\bm r,t)
 \label{noncondensatedensity}
\end{align} 
is the noncondensate number density. 
Here, $U(\bm r,t)=U_{\rm ext}(\bm r,t)+2g[n_c(\bm r,t)+\tilde n(\bm r,t)]$ is the time-dependent effective potential including the external potential $U_{\rm ext}(\bm r,t)$ and the self-consistent Hartree-Fock mean-field potential.
The two collision terms are given by
\begin{align}
	C_{12}[f(1),\Psi]=&\frac{4\pi g^2n_c}{\hbar}
	\int \frac{d\bm p_2}{(2\pi\hbar)^3}\int d\bm p_3\int d\bm       p_4\notag\\
	&\times\delta(m\bm v_c+\bm p_2-\bm p_3-\bm p_4)\notag\\
	&\times\delta(\varepsilon_c+\tilde{\varepsilon}(2)
	-\tilde{\varepsilon}(3)-\tilde{\varepsilon}(4))\notag\\
	&\times[\delta(\bm p_1-\bm p_2)-\delta(\bm p_1-\bm p_3)-       \delta(\bm p_1-\bm p_4)
     ]\notag\\
	&\times\{[1+f(2)]f(3)f(4)\notag\\
 &-f(2)[1+f(3)][1+f(4)]\},\\
	C_{22}[f(1)]=&\frac{4\pi g^2}{\hbar}	
	\int \frac{d\bm p_2}{(2\pi\hbar)^3}\int\frac{d\bm p_3}{(2\pi\hbar)^3}\int d\bm p_4\notag\\
	&\times\delta(\bm p_1+\bm p_2-\bm p_3-\bm p_4)\notag\\
	&\times\delta(\tilde{\varepsilon}(1)+\tilde{\varepsilon}(2)
	-\tilde{\varepsilon}(3)-\tilde{\varepsilon}(4))\notag\\
	&\times\{(1+f(1))(1+f(2))f(3)f(4)\notag\\
        &-f(1)f(2)(1+f(3))(1+f(4))\},
\end{align}
where we have introduced the simplified notation for the distribution function $f(i)=f(\bm r_i,\bm p_i,t)$ with $i = 1,2,3,4$. 
The local energies of the condensate and noncondensate atoms are given by $\varepsilon_c(\bm r,t)=mv^2_c(\bm r,t)/2+\mu_c(\bm r,t)$ and $\tilde\varepsilon(i)=p^2_i/2m+U(\bm r,t)$, 
with the local condensate chemical potential $\mu_c(\bm r,t)$, and the condensate velocity $\bm v_c(\bm r,t)$. 
The dissipation term $R(\bm r,t)$ and the source term $\Gamma_{12}[f,\Psi]$ are given by 
\begin{align}
R(\bm r,t)\equiv & \frac{\hbar\Gamma_{12}[f,\Psi(\bm r,t)]}{2n_c(\bm r,t)},\\
\Gamma_{12}[f,\Psi(\bm r,t)]
\equiv & \int\frac{d\bm p}{(2\pi\hbar)^3}C_{12}\bqty{f(\bm p,\bm r,t),\Psi(\bm r,t)}.\label{source_term}
\end{align}	

In terms of the phase and the amplitude $\Psi(\bm r,t)=\sqrt{n_c(\bm r,t)}e^{i\theta(\bm r,t)}$, we can rewrite the GGPE in terms of the density and velocity $\bm v_c(\bm r,t)=\hbar\nabla \theta(\bm r,t)/m$, given by 
\begin{align}
&\pdv{n_c(\bm r,t)}{t}+\nabla\cdot\bqty{n_c(\bm r,t)\bm v_c(\bm r,t)}
= -\Gamma_{12}[f,\Psi(\bm r,t)],\\
&m\biggl[\pdv{\bm v_c(\bm r,t)}{t}+\frac{1}{2}\nabla v^2_c(\bm r,t)\biggr]
= -\nabla[\mu_c(\bm r,t)+U_{\rm ext}],
\end{align}	
where the local condensate chemical potential $\mu_c(\bm r,t)$ is given by 
\begin{align}
\mu_c(\bm r,t)=&
-\frac{\hbar^2\nabla^2\sqrt{n_c(\bm r,t)}}{2m\sqrt{n_c(\bm r,t)}}
+gn_c(\bm r,t)+2g\tilde{n}(\bm r,t)
\label{mu}.
\end{align} 
In this paper, we are interested in the collective mode in a uniform system.

%%%%%%%%%%%%%%%%%%%%%%%%%%%%%%%%%%%%%%%%%%%%%%%%%%%%%%%%%%%%%%%%%%%%%%%%%%%%%%%%
Following Refs.~\cite{watabeosawanikuni,yellowbook}, we linearize the ZNG equation by writing the physical quantity as $A(\bm r,t)=A_0+\delta A(\bm r,t)$, where $A_0$ is the equilibrium solution and $\delta A(\bm r,t)=\delta A e^{i(\bm q\cdot\bm r-\omega t)}$ is a small fluctuation with a plane-wave solution. Keeping to the first order in $\delta A$, we obtain the following linearized hydrodynamic equations:
\begin{align}
\omega\delta n_c(\bm q,\omega)
= & n_{c0}q\delta v_c(\bm q,\omega)-i\delta\Gamma_{12}(\bm q,\omega),
\label{hydro_continuity}
\\
m\omega\delta v_c(\bm q,\omega)
= & \frac{\hbar^2q^3\delta n_c(\bm q,\omega)}
{4mn_{c0}}
+  gq\delta n_c(\bm q,\omega)\notag\\
&+2qg\delta\tilde{n}(\bm q,\omega)
+qU_{\rm ext}(\bm q,\omega).
\label{hydro_Euler}
\end{align}

For the noncondensate distribution function, we write $f(\bm p,\bm r,t)=f^0(\bm p)+f^0(\bm p)[1+f^0(\bm p)]\nu(\bm p,\bm r,t)$, where  $f^0$ is the equilibrium distribution function in a uniform system 
\footnote{
This form is very useful for linearizing the Boltzmann equation for Bose gases and is often used in the literature.
The factor $f_0(\bm p)(1+f_0(\bm p))$ is motivated by the fact that the expression of the equilibrium Bose distribution function 
$f_0(\bm p)=[z^{-1}_0e^{\beta_0 \tilde\varepsilon_0(\bm p)}-1]^{-1}$) for small deviations of the single particle energy $\tilde \varepsilon_p\to\tilde \varepsilon_p+\delta\tilde \varepsilon_p$ yields $f_0(\bm p)\to f_0(\bm p) + ({\partial f_0(\bm p)}/{\partial\tilde  \varepsilon_0})\delta\tilde\varepsilon_p=-\beta_0f_0(\bm p)(1+f_0(\bm p))\delta\varepsilon_p$.
For more detailed calculations, see, for example, Ref.~\cite{yellowbook}.
}
and $\nu(\bm p,\bm r,t)=\nu(\bm p,\bm q,\omega)e^{i(\bm q\cdot\bm r-\omega t)}$ describes the deviation from the equilibrium distribution with the assumption of the plane-wave solution. 
By substituting this form into the Boltzmann equation, we obtain the linearized Boltzmann equation:
\begin{align}
&i[1+f^0(\bm p)]f^0(\bm p)\Bigl[\Bigl(-\omega+\frac{\bm p\cdot\bm q}{m}\Bigr)\nu(\bm p,\bm q,\omega)\notag\\
&+\beta_0
\frac{\bm p\cdot\bm q}{m}\bqty{2g\delta n(\bm q,\omega)+U_{\rm ext}(\bm q,\omega)}\Bigr]\notag \\
=&-\beta_0[\delta\mu_c(\bm q,\omega)-2g\delta n(\bm q,\omega)]\mathcal{L}_{12}[1]+\mathcal{L}[\nu(\bm p,\bm q,\omega)],
\label{linearizedBoltzmann}
\end{align}
where $\delta n=\delta n_c+\delta \tilde n$ and the linearized collisional operator $\mathcal{L}[\nu]=\mathcal{L}_{12}[\nu]+\mathcal{L}_{22}[\nu]$ is given by 
\begin{align}
    \mathcal{L}_{12}[\nu(1)]=&-\int \frac{d\bm p_2}{(2\pi\hbar)^3}\int d\bm p_3\int d\bm p_4W_{12}(1,2,3,4)\notag\\
    &\times
    [\nu(2)-\nu(3)-\nu(4)], \\
    \mathcal{L}_{22}[\nu(1)]=&-\int \frac{d\bm p_2}{(2\pi\hbar)^3}
    \int \frac{d\bm p_3}{(2\pi\hbar)^3}\int d\bm p_4W_{22}(1,2,3,4)\notag\\
    &\times
    [\nu(1)+\nu(2)-\nu(3)-\nu(4)]. 
\end{align}	
Here, kernels $W_{12}$ and $W_{22}$ are defined by
\begin{align}
W_{12}
=&
\frac{4\pi g^2n_{c0}}{\hbar}[1+f^0(2)]f^0(3)f^0(4)\notag\\
&\times\delta(\bm p_2-\bm p_3-\bm p_4)
\notag\\
&\times\delta(\mu_{c0}+\tilde{\varepsilon}^0(2)
-\tilde{\varepsilon}^0(3)-\tilde{\varepsilon}^0(4))\notag \\
& \times 
\bqty{\delta(\bm p_1-\bm p_2)-\delta(\bm p_1-\bm p_3)-\delta(\bm p_1-\bm p_4)}, 
\\
%\end{align}
%\begin{align}
W_{22}=&\frac{4\pi g^2}{\hbar}
\delta(\bm p_1+\bm p_2-\bm p_3-\bm p_4)\notag \\
&\times\delta(\tilde{\varepsilon}^0(1)+\tilde{\varepsilon}^0(2)
-\tilde{\varepsilon}^0(3)-\tilde{\varepsilon}^0(4))\notag \\
&\times f^0(1)f^0(2)(1+f^0(3))(1+f^0(4)). 
\end{align}	
%and satisfy the following symmetry
%\begin{align}
%	&W_{22}(1,2,3,4)=W_{22}(2,1,3,4)=W_{22}(1,2,4,3)=W_{22}(2,1,4,3),\\
%	&W_{12}(1,2,3,4)=	W_{12}(1,2,4,3).
%\end{align}
We expand $\nu(\bm p)$ and collisional operators $\mathcal{L}_{\alpha}(\bm p)$ for $\alpha=\{12,22\}$ with the spherical harmonics $Y_{l,m}(\theta,\phi)$, resulting in (see Appendix.~\ref{appendix:expansion_of_collision_term} for a detailed discussion on the expansion of $\mathcal{L}_{\alpha}$)
\begin{align}
    \nu(\bm p)
    =&\sum^\infty_{l=0}\sum^l_{m=-l}\sqrt{\frac{4\pi}{2l+1}}\nu^m_l(p)Y_{l,m}(\theta,\phi),
    \label{expandbysphericalharmonics}
    \\
    \mathcal{L}_{\alpha}(\bm p)
    =&\sum^\infty_{l=0}\sum^l_{m=-l}\sqrt{\frac{4\pi}{2l+1}}\mathcal{L}^{ml}_{\alpha}(p)Y_{l,m}(\theta,\phi).
    \label{collisionterm_expandbyYlm}
\end{align}

Let us define the moment $\expval{p^n\nu_l}$ and the temperature dependent function $W_n$ as follows:
\begin{align}
	\expval{p^n\nu_l}\equiv&\int\frac{d\bm p}{(2\pi\hbar)^3}p^n\nu_l(p)f^0(p)[1+f^0(p)],
 \label{defmoment}\\
	W_n=&\int\frac{d\bm p}{(2\pi\hbar)^3}p^nf^0(p)[1+f^0(p)].
\end{align}
One can show that the noncondensate density fluctuation is directly related to the zeroth order moment $\delta\tilde n=\expval{\nu_0}$. 
Multiplying $p^{l+2k}Y^*_{l,m}(\hat{\bm p})$ with the linearized Boltzmann equation Eq.~\eqref{linearizedBoltzmann} and integrating it over the momentum, one obtains the moment equation:
\begin{align}
	&-i\omega\expval{p^{l+2k}\nu_l}+i\frac{q}{m}\frac{l}{2l-1}\expval{p^{l+2k+1}\nu_{l-1}}\notag \\
	&+i\frac{q}{m}\frac{l+1}{2l+3}\expval{p^{l+2k+1}\nu_{l+1}}\notag\\
&+i\beta\frac{q}{m}W_{l+2k+1}(2g\delta n_c+2g\expval{\nu_0}+U_{\rm ext})\delta_{l,1}\notag\\
=&\dv{t}\expval{p^{l+2k}\nu_l}_{\rm coll}. 
\label{not_closed_moment_eq}
\end{align}
The collision term on the right-hand side is given by
\begin{align}
	\dv{t}\expval{p^{l+2k}\nu_l}_{\rm coll}
	=&-J_{l}[p^{l+2k},\nu_{l}(p)]\notag\\
 &+\beta_0\pqty{\frac{\hbar^2q^2}{4mn_{c0}}-g}J_{0,12}[p^{2k},1]\delta n_c\delta_{l,0},
 \label{rhsofmomenteq}
\end{align}
where we defined the total collision integral as 
\begin{align}
	J_l[p^n,\nu_l]=J_{l,12}[p^n,\nu_{l}(p)]+J_{l,22}[p^n,\nu_{l}(p)],
\end{align}
and the collision integral $J_{l,\alpha}$ for $\alpha={12,22}$ by
\begin{align}
J_{l,\alpha}[p^n,\nu_{l}(p)]=-\int\frac{d\bm p}{(2\pi\hbar)^3}p^n\mathcal{L}^l_\alpha[\nu_l(p)].
\end{align}
So far, the moment equation in Eq.~\eqref{not_closed_moment_eq} is the exact consequence obtained from the linearized ZNG equations. 

To solve the moment equation, we must express the collision term $J_{l,\alpha}$ in terms of the moments.
For this purpose, we expand $\nu_l(p)$ in power series of $p$:
\begin{align}
	\nu_l(p)=\sum_{k=0}C^l_kp^{l+2k}, \label{expandfluctuationbySonine}
\end{align}
where, in the classical kinetic theory, the coefficient $C^l_k$ is given by the Sonine or Hermite polynomials depending on the system coordinates \cite{sonine_polynomial}.
By substituting the expression in Eq.~\eqref{expandfluctuationbySonine} into the definition of the moment in Eq.~\eqref{defmoment}, we obtain the relation between the moment $\expval{p^{l+2k}\nu_l}$ and  the expansion coefficient $C^l_{k'}$ given in the form
\begin{align}
	\expval{p^{l+2k}\nu_l}
	=\sum_{k'}C^l_{k'}W_{2l+2k+2k'}.
 \label{momentintermofsoninepolynomial}
\end{align} 

\begin{figure}[tbp]
	% \centering
	% \includegraphics[width=\linewidth]{Figs/relaxation_rates.pdf}
 %\includegraphics[width=\linewidth]{Figs/relaxation_rates_Modified.pdf}
  \includegraphics[width=\linewidth]{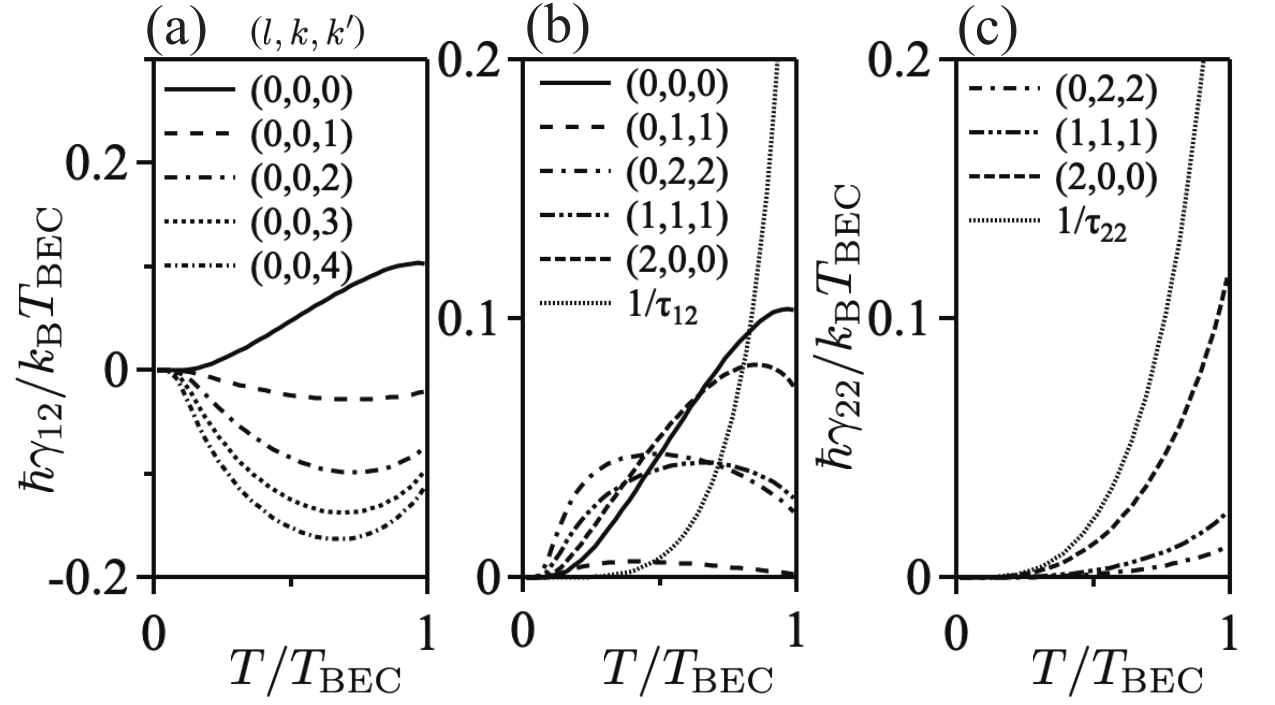}
	\caption{
Collision rates appearing in the moment equation. 
 The plot label is presented as $(l,k,k^\prime)$, e.g. $(l,k,k^\prime)=(0,0,0)$. 
Panels (a) and (b) show the collision rates originating from the $C_{12}$ collision process. 
Panel (c) shows the collision rates originating from the $C_{22}$ collision process.
We plot the inverse relaxation time $1/\tau^0_{12}$ and  $1/\tau^0_{22}$. 
Here, the interaction strength is $gn_{0}=0.3k_{\rm B}T_{\rm BEC}$.
 }
 \label{fig:relax_rates}
\end{figure}

Similarly, the collision integral can be rewritten in terms of the expansion coefficient $C^l_{k'}$, given by 
\begin{align}
	J_{l}[p^{l+2k},\nu_{l}(p)]
	=&\sum_{\alpha}\sum_{k'}C^l_{k'}J_{l,\alpha}[p^{l+2k},p^{l+2k'}]\notag\\
	\equiv&
 \sum_{\alpha}\sum_{k'}C^l_{k'}\gamma^l_{\alpha,kk'}W_{2l+2k+2k'}. 
 \label{collision_integral_by_relaxation_rate}
\end{align}
Here, we defined the {\it generalized collision rate} $\gamma^l_{\alpha,kk'}$ given by 
\begin{align}
	\gamma^l_{\alpha,kk'}\equiv\frac{1}{W_{2l+2k+2k'}}J_{l,\alpha}[p^{l+2k},p^{l+2k'}].
 \label{def_moment_relaxation_rate}
\end{align}
(see Appendix.\ref{appendix:collisionintegral} for the explicit expression of the collision rate $\gamma^l_{\alpha,kk'}$.)
We note that the generalized collision rate $\gamma^l_{12,kk'}$ can be either positive or negative, depending on the indices $l,k$ and $k'$ (Fig.~\ref{fig:relax_rates} (a)). 

The so-called relaxation-time approximation can be implemented by neglecting the $l,k,$ and $k'$ dependence, where $\gamma^l_{\alpha,kk'}$ is replaced with $\gamma_\alpha$, given by 
\begin{align}
	J_{l}[p^n,\nu_{l}(p)]
	=&\sum_{\alpha}\gamma_{\alpha}\sum_{k'}W_{2l+2k+2k'}C^l_{k'}\notag\\
	=&\sum_{\alpha}\gamma_{\alpha}\expval{p^{l+2k}\nu_l}. 
 \label{appendix:qualitative_relaxation_time_approximation}
\end{align} 
The approach of introducing the single collision rate has often been employed in normal systems \cite{watabenikuni,watabeosawanikuni,asano}.
In Fig.~\ref{fig:relax_rates}, we plot the collision rate defined by Eq.~\eqref{def_moment_relaxation_rate}, and also the inverse relaxation time $1/\tau_{\alpha}$ associated with the $C_{\alpha}$ collision process (See the appendix~\ref{appendix:collisionintegral}).
The collision rates $\gamma^l_{22,kk'}$ and $1/\tau_{22}$ for the collision process between noncondensate atoms have similar monotonically increasing temperature dependence (see Fig.~\ref{fig:relax_rates} (c)). 

As in Figs.~\ref{fig:relax_rates} (a) and (b), due to the divergent property of the Bose distribution function, the collision rate $\gamma^l_{12,kk'}$ has a significant temperature dependence for $l=0,1$. 
This is the striking difference between Bose-condensed systems and normal systems. 
As in the collision rate $\gamma^l_{22,kk'}$ in Eq.~\eqref{appendix:I_22}, the integrand has a form of the distribution function multiplied by the power function of the single particle momentum, which suppresses the divergent behavior of the Bose distribution function for a large value of $l,k$, and $k'$. Thus, one can replace the collision rates $\gamma^l_{\alpha,kk'}$ with a dominant collision rate. 
Indeed, we numerically confirmed that the collision rate in the moment equation for $l>2$ has a temperature dependence similar to that of $l=2$.

To make use of this fact, we systematically develop the relaxation-time approximation for the moment equation of the Bose-condensed gas (see Appendix \ref{appendix:relaxationtimeapprox}). 
By applying the relaxation-time approximation to Eq.~\eqref{collision_integral_by_relaxation_rate} and using the relation Eq.~\eqref{momentintermofsoninepolynomial}, we obtain the following moment equation:
\begin{align}
 \omega\expval{\nu_0}=&\frac{q}{3m}\expval{p\nu_{1}}
	+i\beta\pqty{\frac{\hbar^2q^2}{4mn_{c0}}-g}\gamma^0_{12,00}W_{0}\delta n_c\notag \\
	&-i\biggl[\pqty{\gamma^0_{12,00}-\frac{W^2_2}{D}(\gamma^0_{12,01}-\gamma^0_{12,00})}\expval{\nu_0}\notag\\
 &+\frac{W_2W_0}{D}(\gamma^0_{12,01}-\gamma^0_{12,00})\expval{p^2\nu_0}
 \biggr], 
\label{momenteq_l0k0}
\\ 
	\omega\expval{p\nu_1}=&\frac{q}{m}\expval{p^{2}\nu_{0}}
		+\frac{2q}{5m}\expval{p^{2}\nu_{2}}\notag\\
  &+\beta_0\frac{q}{m}W_{2}[2g\delta n_c+2g\expval{\nu_0}+U_{\rm ext}], 
  \label{momenteq_l1k0}
  \\ 
	\omega\expval{p^{2}\nu_0}=&\frac{q}{3m}\expval{p^{3}\nu_{1}}
	+i\beta_0\pqty{\frac{\hbar^2q^2}{4mn_{c0}}-g}\gamma^0_{12,01}W_{2}\delta n_c\notag \\
	&-i\biggl[\frac{W_2W_4}{D}\pqty{\gamma^0_{12,01}-\gamma^0_{12,11}}\expval{\nu_0}\notag\\
 &+\pqty{\frac{W_4W_0}{D}\gamma^0_{12,11}-\frac{W^2_2}{D}\gamma^0_{12,01}}\expval{p^2\nu_0}\biggr], 
 \label{momenteq_l0k1}
 \\ 
	\omega\expval{p^{2k}\nu_0}=&\frac{q}{3m}\expval{p^{2k+1}\nu_{1}}\notag\\
	&+i\beta\pqty{\frac{\hbar^2q^2}{4mn_{c0}}-g}\gamma^0_{12,0k}W_{2k}\delta n_c\notag \\
		&-i\biggl[(\gamma^0_{12,0k}-\gamma^0_{12,22}-\gamma^0_{22,22})\frac{W_{2k}W_4}{D}\notag\\
  &-(\gamma^0_{12,1k}-\gamma^0_{12,22}-\gamma^0_{22,22})\frac{W_{2k+2}W_2}{D}\biggr]\expval{\nu_0}\notag \\
	&-i\biggl[(\gamma^0_{1k,12}-\gamma^0_{12,22}-\gamma^0_{22,22})\frac{W_{2k+2}W_0}{D}\notag\\
	&-(\gamma^0_{12,0k}-\gamma^0_{12,22}-\gamma^0_{22,22})\frac{W_{2k}W_2}{D}\biggr]
	\expval{p^2\nu_0}\notag \\
	&-i[\gamma^0_{12,22}+\gamma^0_{22,22}]\expval{p^{2k}\nu_0}, 
 \label{momenteq_l0k}
\\ 
	\omega\expval{p^{2k+1}\nu_1}=&\frac{q}{m}\expval{p^{2k+2}\nu_{0}}
		+\frac{2q}{5m}\expval{p^{2k+2}\nu_{2}}\notag\\
  &+\beta\frac{q}{m}W_{2k+2}[2g\delta n_c+2g\expval{\nu_0}+U_{\rm ext}]\notag \\
	&-i\biggl[-[\gamma^1_{12,11}+\gamma^1_{22,11}]\frac{W_{2+2k}}{W_2}\expval{p\nu_1}\notag \\
	&+[\gamma^1_{12,11}+\gamma^1_{22,11}]\expval{p^{1+2k}\nu_1}\biggr], 
 \label{momenteq_l1k}
\\ 
	\omega\expval{p^{l+2k}\nu_l}=&\frac{q}{m}\frac{l}{2l-1}\expval{p^{l+2k+1}_1\nu_{l-1}}\notag \\
		&+\frac{q}{m}\frac{l+1}{2l+3}\expval{p^{l+2k+1}_1\nu_{l+1}}\notag \\
  &-i(\gamma^2_{12,00}+\gamma^2_{22,00})\expval{p^{l+2k}\nu_l}, 
  \label{momenteq_l2}
\end{align}
where $D=W_0W_4-W^2_2$.
Equations~\eqref{momenteq_l0k0}-\eqref{momenteq_l2} can be derived from the moment equation of order $(l=0,k=0)$, $(l=1,k=0)$, $(l=0,k=1)$, $(l=0,k\geq2)$, $(l=1,k\neq0)$, and $(l\geq2)$, respectively. 
The condensate equations in Eqs.~\eqref{hydro_continuity} and \eqref{hydro_Euler} can be expressed in terms of the moments as
\begin{align}
\omega\delta n_c
=&n_{c0}q\delta v_c-i\beta\pqty{\frac{\hbar^2q^2}
{4mn_{c0}}-g}\gamma^0_{12,00}W_{0}\delta n_c\notag \\
 &+i\bqty{\gamma^0_{12,00}-\frac{W^2_2}{D}(\gamma^0_{12,01}-\gamma^0_{12,00})}\expval{\nu_0}\notag\\
 &+i\bqty{\frac{W_2W_0}{D}(\gamma^0_{12,01}-\gamma^0_{12,00})}\expval{p^2\nu_0}, 
 \label{moment_nc}
 \\
m\omega\delta v_c
=&\frac{\hbar^2q^3\delta n_c}
{4mn_{c0}}+gq\delta n_c+2gq\delta\tilde{n}
+qU_{\rm ext}. 
\label{moment_euler}
\end{align}
We emphasize that the relaxation-time approximation introduced above satisfies the required conservation laws for the number of particles, momentum, and local energy~\cite{energyfunctional}. 

In short, using the relaxation-time approximation, we developed the ZNG moment equations as in Eqs.\eqref{momenteq_l0k0}-\eqref{momenteq_l2}. 
Our moment equations generalize previous studies for normal gases \cite{watabeosawanikuni,asano} to Bose-condensed gases, which are more complicated because of the coupling to the condensate component.
We note that the ZNG moment equations are not closed due to their hierarchical nature. Therefore, in order to solve them, we need to truncate at a sufficiently high moment term. 
In the next section, we discuss the connection between the Landau two-fluid equation and the ZNG moment equation in detail.

\section{Reduction to the Landau two-fluid equation}\label{sect:two-fluid} 

By taking the hydrodynamic limit, we show that the ZNG moment equations can be reduced to the Landau two-fluid equations including transport coefficients. 
The derivation presented in this section has a close resemblance to the standard Chapman--Enskog approach~\cite{yellowbook,chapmannenskog}, which expands the distribution function around the local equilibrium distribution function $f^{(0)}$. 
In the framework of the moment method, we expand the Boltzmann equation by the dimensionless parameter $\omega/\gamma^l_{\alpha,kk'}$. 
A similar approach is presented in the case of the two-component Fermi gas by using the linearized Boltzmann equation~\cite{watabeosawanikuni,narushima}. 

To obtain the two-fluid equations, one needs to relate the moments to the hydrodynamic variables. 
Indeed, using the local equilibrium solution, we can obtain the following relations (see Appendix~\ref{appendix:leq} for details) 
\begin{align}
    \expval{\nu_0}=\delta\tilde n,
    \;\;\;
    \expval{p\nu_1}=3m\tilde n\delta v_n,
    \;\;\;
    \expval{p^2\nu_0}=3m\delta\tilde P, 
    \label{momentsbydensityvelocitypressurefluctuation}
\end{align}
where the moments  $\expval{\nu_0}$, $\expval{p\nu_1}$, and $\expval{p^2\nu_0}$ are proportional to the fluctuations of the noncondensate density $\delta\tilde n$, velocity field $\delta v_n$, and pressure $\delta \tilde P$. 

To derive the two-fluid equations, we introduce the Thomas--Fermi approximation \cite{yellowbook}.
For notational conciseness, we omit the sub(super)-script zero in the equilibrium thermodynamic quantities.
Equations~\eqref{momenteq_l0k0}, \eqref{momenteq_l0k1}, and \eqref{momenteq_l1k0} can be rewritten in terms of the above hydrodynamic quantities as
\begin{align}
\omega\delta\tilde n=&
		q\tilde n\delta v_n
	-i\frac{\beta_0 gn_{c}}{\tau_{12}}\delta n_c-i\frac{\sigma_{2}n_{c}}{\tau_{12}\tilde n}\delta\tilde n\notag\\
 &-i\frac{\sigma_{1}\beta n_{c}}{\tau_{12}\tilde n}\delta\tilde P,
 \label{landauzngtwofluidcontinuity}
 \\
	\omega\delta v_n=&\frac{q}{m\tilde n}\delta\tilde P
		+\frac{2q}{15m^2\tilde n}\expval{p^{2}\nu_{2}}\notag\\
	&+\frac{q}{m}(2g\delta n_c+2g\delta \tilde n+U_{\rm ext}),
 \label{momentl1k0coursegrainedvar}
 \\
 	\omega\delta\tilde P=&
		\frac{q}{9m^2}\expval{p^{3}\nu_{1}}
	+i\frac{2\beta g^2n^2_{c}}{3\tau_{12}}\delta n_c\notag\\
        &
        +i\frac{2gn^2_{c}\sigma_{2}}{3\tilde n\tau_{12}}\delta\tilde n+i\frac{2\beta gn^2_{c}\sigma_{1}}{3\tilde n\tau_{12}}\delta \tilde P,
        \label{momentl0k1coursegrainedvar}
\end{align}
 where we used the relation 
 % $\gamma^0_{12,01}=-2mgn_{c}(W_0/W_2)\gamma^0_{12,00}$, $\gamma^0_{12,11}=(2mgn_{c})^2(W_0/W_4)\gamma^0_{12,00}$ and $\gamma^0_{12,00}=n_{c}/(W_0\tau_{12})$
 between collision rates
 Eq.~\eqref{appendix:gamma000_tau}-\eqref{appendix:gamma012_gamma002}
(see Appendix~\ref{appendix:collisionintegral}). 
We also defined the hydrodynamic coefficients
\begin{align}
 &   \sigma_{1}=-\frac{W_2W_0}{D} \biggl ( 2mgn_{c}+\frac{W_2}{W_0} \biggr ),
 \label{sig1}
    \\
& 
    \sigma_{2}
	=\frac{\beta W_2}{3D} \biggl ( \frac{W_4}{m}+2gn_cW_2 \biggr ), 
 \label{sig2}
\end{align}
with the use of the relation 
\begin{align}
	&\tilde n=\frac{W_2}{3mk_BT}. 
\end{align}
By writing the coefficient $W_n$ in terms of the Bose-Einstein function $g_n(z)=\sum_{l=1} z^l/l^n $ using Eq.~\eqref{appendix:Wn_by_gn}, one can show that the above definition is consistent with the hydrodynamic coefficients discussed by ZNG \cite{ZNG_original,yellowbook,chapmannenskog}. 

Due to the hierarchical structure of the moment equation, the equations \eqref{landauzngtwofluidcontinuity}, \eqref{momentl1k0coursegrainedvar} and  \eqref{momentl0k1coursegrainedvar} are not closed, which are coupled to the higher order of the moments through the terms $\expval{p^{2}\nu_{2}}$ or $\expval{p^{3}\nu_{1}}$.
Taking the hydrodynamic limit $\omega/\gamma^l_{\alpha,kk'}\to0$, we can relate $\expval{p^2\nu_2}$ and $\expval{p^3\nu_1}$ with the hydrodynamic variables $\delta \tilde n$, $\delta v_n$, and $\delta \tilde P$ as (see the Appendix~\ref{appendix:two-fluid}) 
\begin{align}
	\expval{p^{2}\nu_2}=&-10im\eta q\delta v_n,
 \label{p2nu2momentbyviscosityandvelocity}\\
 	\expval{p^{3}\nu_1}=&
 15m^2\tilde P\delta v_n+i\frac{6q\sigma_{4}m^2T\kappa}{\tilde n}\delta\tilde n\notag\\
	&-i\frac{6q\sigma_{3}m^2T\kappa}{\tilde P}\delta\tilde P,
 \label{p3nu1momentbykappa}
\end{align}
where the pressure is given by
\begin{align}
        &\tilde P
	=\frac{W_4}{15m^2k_BT}.
\end{align}
We also defined dimensionless hydrodynamic coefficients
\begin{align}
    &\sigma_3=\frac{2W_4W_0}{5D},\label{sig3}
    \\
    &\sigma_4=\frac{2W^2_2}{3D},\label{sig4}
\end{align}
 and the shear viscosity $\eta$ and the thermal conductivity $\kappa$ given by 
\begin{align}
    \eta=&\tau_{\eta}\frac{\beta_0W_4}{15m^2}.\label{viscosity}\\
    \kappa=& \tau_\kappa\frac{k_B\beta^2}{12m^4}\pqty{W_6-\frac{W^2_4}{W_2}},
\end{align}
where $\tau_\eta=(\gamma^2_{12,00}+\gamma^2_{22,00})^{-1}$ and $\tau_\kappa=(\gamma^1_{12,11}+\gamma^1_{22,11})^{-1}$ are the relaxation time associated with the shear viscosity and thermal conductivity, respectively. 
We note that the detailed expression of $\tau_{\eta}$ and $\tau_{\kappa}$ slightly differs from the ones derived by the Chapman--Enskog methods~\cite{chapmannenskog} (see Appendix.~\ref{appendix:tau_kappaeta} for the discussion in detail).

By substituting the first order correction shown in Eqs.~\eqref{p2nu2momentbyviscosityandvelocity} and \eqref{p3nu1momentbykappa} into Eqs.~\eqref{momentl1k0coursegrainedvar} and \eqref{momentl0k1coursegrainedvar}, we obtain the two-fluid equations including transport coefficients, which are given by 
\begin{align}
	\omega\delta n_c
 	=&n_{c}q\delta v_c
	+i\frac{\beta gn_{c}}{\tau_{12}}\delta n_c+i\frac{\sigma_{2}n_{c}}{\tau_{12}\tilde n_0}\delta\tilde n+i\frac{\sigma_{1}\beta n_{c}}{\tilde n\tau_{12}}\delta\tilde P, 
 \label{landautwofluid_nc}
 \\
	\omega\delta v_c
	=&\frac{gq}{m}\delta n_c+\frac{2gq}{m}\delta\tilde{n}
	+\frac{q}{m}U_{\rm ext}, 
\label{landautwofluid_vc}
 \\
	\omega\delta\tilde n=&
		q\tilde n\delta v_n
	-i\frac{\beta gn_{c}}{\tau_{12}}\delta n_c-i\frac{\sigma_{2}n_{c}}{\tau_{12}\tilde n}\delta\tilde n-i\frac{\sigma_{1}\beta n_{c}}{\tilde n\tau_{12}}\delta\tilde P, 
 \label{landautwofluid_nn}
 \\
	\omega\delta v_n
		=&\frac{q}{m\tilde n}\delta \tilde P
		+\frac{q}{m}(2g\delta n_c+2g\delta\tilde n+U_{\rm ext})-i
	\frac{4q^2\eta}{3m\tilde n} \delta v_n, 
 \label{landautwofluid_vn}
 \\
	\omega\delta \tilde P	=&
		\frac{5\tilde P}{3}q\delta v_n
	+i\frac{2\beta g^2 n^2_{c}}{3\tau_{12}}\delta n_c
\notag \\	
 & +\frac{2i}{3}\pqty{\frac{\sigma_{2}gn^2_{c}}{\tau_{12}\tilde n}
 +\frac{q^2\sigma_{4}T\kappa}{\tilde n}}\delta\tilde n\notag\\
	&+\frac{2i}{3}\pqty{\frac{\sigma_{1}\beta g n^2_{c}}{\tilde n\tau_{12}}
 -\frac{q^2\sigma_{3}T\kappa}{\tilde P}}\delta \tilde P. 
 \label{landautwofluid_P}
\end{align}
The two-fluid equations in the above form were first introduced in Refs.~\cite{ZNG_original,comparerelaxationrates}.
Since the above two-fluid equation includes the dissipation from the relaxation time $\tau_{12}$ and the transport coefficients $\eta$ and $\kappa$, we shall call Eq.~\eqref{landautwofluid_nc}-\eqref{landautwofluid_P} {\it dissipative Landau two-fluid equations}.

We now discuss the eigenmodes (solutions in the case $U_{\rm ext}=0$) in the dissipationless limit $\tau_{12},\kappa,\eta\to0$.
We first notice that terms associated with $\tau_{12}$ in Eqs.~\eqref{landautwofluid_nc}, \eqref{landautwofluid_nn}, and \eqref{landautwofluid_P} are originating from the linearized source term:
\begin{align}
    \delta\Gamma_{12} = -\frac{\beta gn_{c}}{\tau_{12}}\delta n_c-\frac{\sigma_{2}n_{c}}{\tau_{12}\tilde n}\delta\tilde n-\frac{\beta n_{c}}{\tilde n\tau_{12}}\sigma_{1}\delta\tilde P.
    \label{sourceterm_physicalquantities}
\end{align}

Introducing the velocity potential in the Fourier space: $\delta v_{c,n}(\bm q,\omega)=iq\phi_{c,n}(\bm q,\omega)$ and 
inserting Eqs.~\eqref{landautwofluid_nc}, \eqref{landautwofluid_nn}, and \eqref{landautwofluid_P} into Eqs.~\eqref{landautwofluid_vc}, and \eqref{landautwofluid_vn}, 
we have
\begin{align}
	m\omega^2\phi_c
	=&gn_{c0}q^2\phi_c+2g\tilde n_0q^2\phi_n+g\delta\Gamma_{12},
 \label{Twofluid_phi_c}
 \\
	m\omega^2\phi_n
		=&\frac{q^2}{\tilde n_0}\frac{5\tilde P}{3}\phi_n
  -\frac{2gn_{c}}{3\tilde n}\delta\Gamma_{12}\notag\\
  &
		+2gq^2(n_{c}\phi_c+\tilde n\phi_n).
  \label{Twofluid_phi_n}
\end{align}
where we have used  Eq.~\eqref{sourceterm_physicalquantities}.
The hydrodynamic limit is also assumed ($\kappa,\eta=0$).
Combining Eq.~\eqref{landautwofluid_nn} and \eqref{landautwofluid_P}, we can express the linearized source term $\delta\Gamma_{12}$ in terms of the velocity potential $\phi_{c,n}$ as 
\begin{align}
	\delta\Gamma_{12}
	=&-\frac{\sigma_{\rm H}n_c}{1-i\omega\tau_{\mu}}\pqty{\phi_c-\frac{2}{3}\phi_n}q^2,
 \label{delmu_by_phic_phin}
\end{align}
where we defined the relaxation time characterizing the diffusive equilibrium between the condensate and noncondensate:
\begin{align}
    \frac{1}{\tau_{\mu}}
    =\frac{\beta g n_c}{\tau_{12}}\pqty{\frac{\frac{5}{2}\tilde P+2g\tilde nn_c+\frac{2}{3}\beta g W_0 gn^2_c}
    {\frac{5}{2}\beta g W_0\tilde P-\frac{3}{2}g\tilde n^2}-1}
    \equiv\frac{\beta g n_c}{\tau_{12}\sigma_{\rm H}}.
    \label{chemicalpotentialrelaxationtime}
\end{align}
Taking Landau limit $\omega\tau_\mu\to0$, one can show that sound velocity $u=\omega/q$ obtained from Eqs.~\eqref{Twofluid_phi_c} and \eqref{Twofluid_phi_n} is given as the solution of
\begin{align}
    &\bqty{u^2-\frac{gn}{m}(1-\sigma_H)}
    \bqty{u^2-\frac{5\tilde P}{3m\tilde n} -\frac{2g\tilde n}{m}
    \pqty{1-\frac{2\sigma_Hn^2_c}{9\tilde n^2}}}\notag\\
    &-\frac{4g^2\tilde n n_c}{m^2}\pqty{1+\frac{\sigma_Hn_c}{3\tilde n}}^2
    =0. 
    \label{extendedzgn}
\end{align}
It has been shown that Eq.~\eqref{extendedzgn} determines the first and second sound velocities in a dilute Bose gas in precise agreement with those determined by the usual Landau two-fluid equations without dissipation \cite{comparerelaxationrates,ZNG_original,griffinzaremba}.
In the next section, we shall compare the results obtained from the ZNG moment equations in Eqs.~\eqref{momenteq_l0k0}-\eqref{momenteq_l2} with those obtained by the dissipative Landau two-fluid hydrodynamics given in 
 Eqs.~\eqref{landautwofluid_nc}-\eqref{landautwofluid_P} and its dissipationless limit given in Eq.~\eqref{extendedzgn}.

% -----------------------[ Result ]------------------------- %	
\section{Denstiy response function}\label{sect:result}

We numerically solve the moment equations in Eqs.~\eqref{momenteq_l0k0}-\eqref{momenteq_l2} and the generalized GP hydrodynamic equations in Eqs.~\eqref{moment_nc} and \eqref{moment_euler} to study the crossover of collective excitations between the hydrodynamic and collisionless regimes.
In the framework of the moment method, the density response function is directly accessible and extremely useful for investigating collective excitations. 
Using the relation $\delta\tilde{n}=\expval{\nu_0}$, one finds that the response function $\chi$ is given by
\begin{align}
\chi(\bm q,\omega)= & \chi_{n_c}(\bm q,\omega)+\chi_{\tilde n}(\bm q,\omega),
\end{align}
where 
\begin{align}
\chi_{n_c}(\bm q,\omega)= & \frac{\delta n_c(\bm q,\omega)}{U_{\rm ext}(\bm q,\omega)}, 
\label{condensateresponse}
\\
\chi_{\tilde n}(\bm q,\omega)= & \frac{\expval{\nu_0}(\bm q,\omega)}{U_{\rm ext}(\bm q,\omega)}. 
\label{noncondensateresponse}
\end{align}
Here, $\chi_{n_c}$ and $\chi_{\tilde n}$  are condensate and noncondensate density response functions, respectively.

In the finite-temperature Bose-condensed gas, there are two distinct regions: hydrodynamic and collisionless regimes.
The first and second sounds emerge in the hydrodynamic limit, whereas in the collisionless regime, the Bogoliubov mode emerges alone~\cite{williamsfiniteTstringari}. 
We are interested in the crossover regime between them and tackle this problem by using the moment method. 
In the following, we take the moment up to $l=k=50$ for the ZNG moment equation.

To numerically solve the moment equations, one must evaluate the equilibrium condensate density.
In equilibrium, the distribution function is given by the Bose--Einstein distribution $f=[z^{-1}\exp(p^2/2mk_{\rm B}T)-1]^{-1}$ with fugacity $z=e^{(\mu-2gn)/k_{\rm B}T}$.
Using this equilibrium distribution function, we self-consistently determine the equilibrium noncondensate density in Eq.~\eqref{noncondensatedensity} and the chemical potential $\mu$.
Within the Hartree--Fock mean-field approximation, the critical temperature  $T_{\rm BEC}$  is known to be identical to that of the ideal Bose gas.

\begin{figure}[tbp]
    \centering
    \includegraphics[scale=.5]{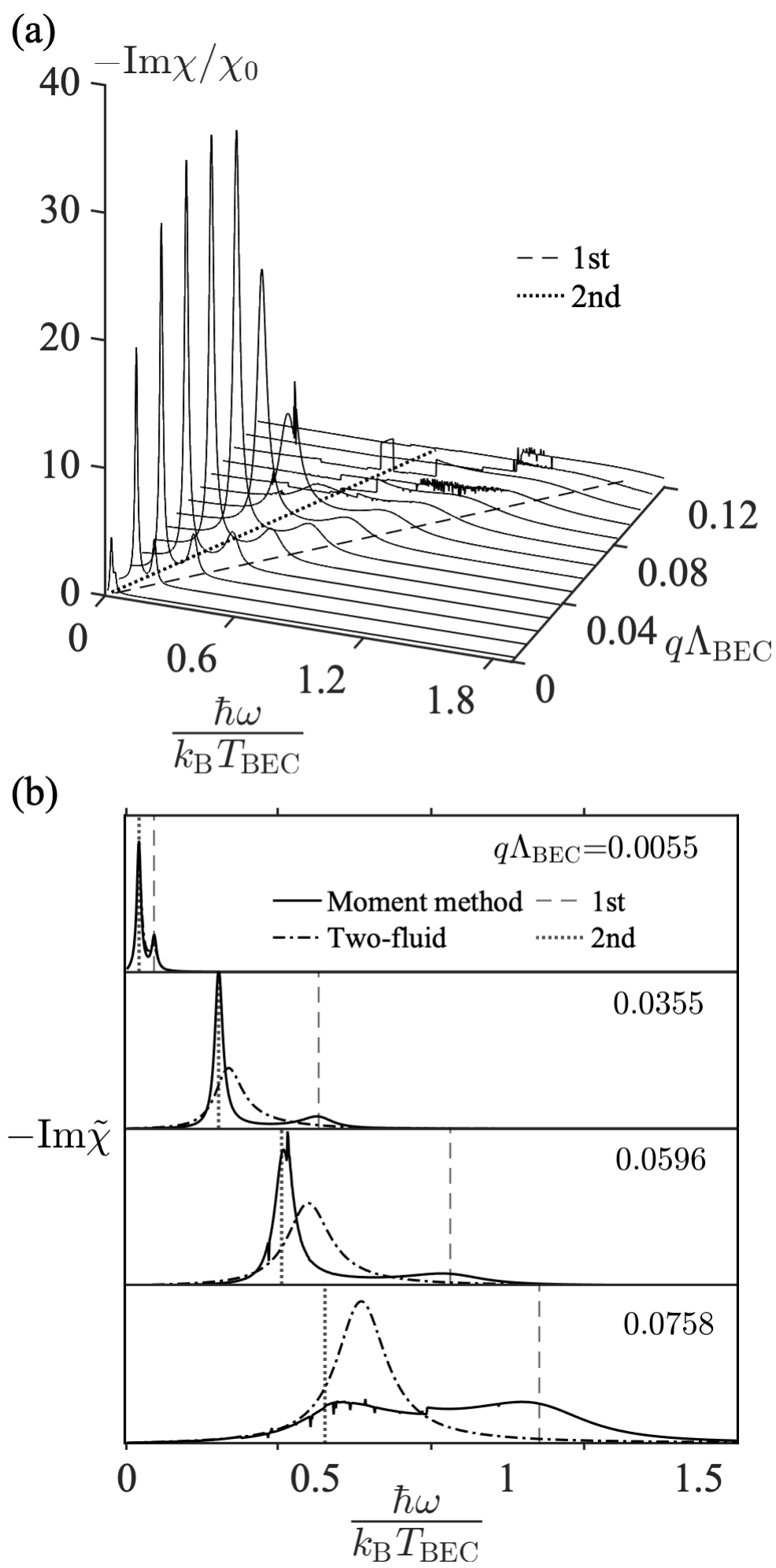}
    \caption{(a) Imaginary part of the density response function $-\Im\chi(\omega,q)$ in the $\omega$-$q$ plane. 
    (b) Imaginary part of the normalized density response function $-\Im\tilde\chi(\omega)$ for several values of the wavenumber $q$. 
    In both panels (a) and (b), the dashed and dotted lines show the dispersion relations of the first and second sounds in the dissipationless Landau limit obtained from Eq.~\eqref{extendedzgn}. 
    The temperature and interaction strength are $T=0.5T_{\rm BEC}$ and $gn_0=0.3k_{\rm B}T_{\rm BEC}$, respectively.
    }
    \label{fig:chiwq2d3d}
\end{figure}

Figure~\ref{fig:chiwq2d3d} (a) shows the imaginary part of the density response function as a function of the frequency $\omega$ and the wavenumber $q$ for $gn_0=0.3k_{\rm B}T_{\rm BEC}$. 
We choose the temperature at $T=0.5T_{\rm BEC}$. 
The density response function is scaled by $\chi_0=n/(k_{\rm B}T_{\rm BEC})$, where $n=\zeta(\tfrac{3}{2})/\Lambda_{\rm BEC}$ with the thermal de Blogie wavelength evaluated at the BEC critical temperature $T_{\rm BEC}$, given by $\Lambda_{\rm BEC}=\sqrt{h/(2\pi mk_{\rm B}T_{\rm BEC})}$. 

By changing the wavenumber $q$, the collisionless regime $\omega\tau\gg1$ and hydrodynamic regime $\omega\tau\ll1$ are achieved, where $\omega$ is the frequency of the collective modes and $\tau$ is a characteristic relaxation time. 
We note that the linearized Boltzmann equation is only valid for the long-wavelength limit $q\ll 1/\Lambda_{\rm BEC}$ \cite{watabeosawanikuni}. 
Even within this limitation, one can address the crossover regime. 
In the small wavenumber region, there are two sharp peaks corresponding to the first and second sounds emerging at the higher and lower frequencies, respectively.
With increasing the wavenumber, the second-sound peak vanishes because of the coupling with incoherent modes.
In contrast, the peak from the first sound becomes significantly broadened, which shows the crossover to the collisionless regime.

To compare the results of the moment method with the hydrodynamic two-fluid theory, we show the imaginary part of the density response function as a function of $\omega$ (Fig.~\ref{fig:chiwq2d3d} (b)). 
Since the peak heights of the two methods are different, we normalize each function by the total weight, given by $\tilde\chi(\omega)=\chi(\omega)/\int d\omega(-\Im\chi)$. 
The two methods reasonably provide almost the same results in the long-wavelength limit $q\Lambda_{\rm BEC}=0.0055$ since the moment method reduces to the two-fluid equations with small transport coefficients. 

However, at $q\Lambda_{\rm BEC}=0.0355$, the dissipative Landau two-fluid equations given by Eqs.~\eqref{landautwofluid_nc}-\eqref{landautwofluid_P} predicts a broader peak for the second sound compared to the moment method, whereas peak position is shifted. 
Moreover, unlike the prediction by the moment method, the first sound is no longer visible in the dissipative Landau two-fluid model. 
This observation for the first sound holds at $q\Lambda_{\rm BEC}=0.0596$. 
As for the second sound, on the other hand, incoherent excitations emerge and violate the response peak of the second sound.

This incoherent excitation significantly affects the second sound mode. 
In the moment method at $q\Lambda_{\rm BEC}=0.0758$, 
the peak height from the second sound becomes lower than the first sound.
% In the moment method, the first sound peak is broadened but still visible around the frequency predicted by the Landau two-fluid equation. 

This observation reveals that the weight of the two modes in the density response function switches in the crossover regime. 
While the first sound peak smoothly decays in the crossover regime, the second sound loses its weight abruptly due to the coupling with incoherent excitations. 
In the collisionless limit, there are neither distinct first nor distinct second sound modes. 
The damping rate obtained by the dissipative Landau two-fluid equation, Eq.~\eqref{landautwofluid_nc}-\eqref{landautwofluid_P} is significantly large, which results in the vanishing first sound peak. 
Moreover, the incoherent excitations, which are important in the crossover regime, cannot be captured by the two-fluid equation as expected.
These are in stark contrast to the moment method. 

In order to address the crossover regime from the hydrodynamic (first or second sound) modes to the collisionless (Bogoliubov) sound mode, we investigate the temperature dependence of the imaginary part of the density response function.
First, we discuss the dimensionless parameter $\omega_{1,2}\tau$, where $\omega_1$ and $\omega_2$ are the first and second sound frequencies, respectively.

\begin{figure}[tbp]
    \centering
      \includegraphics[scale=.4]{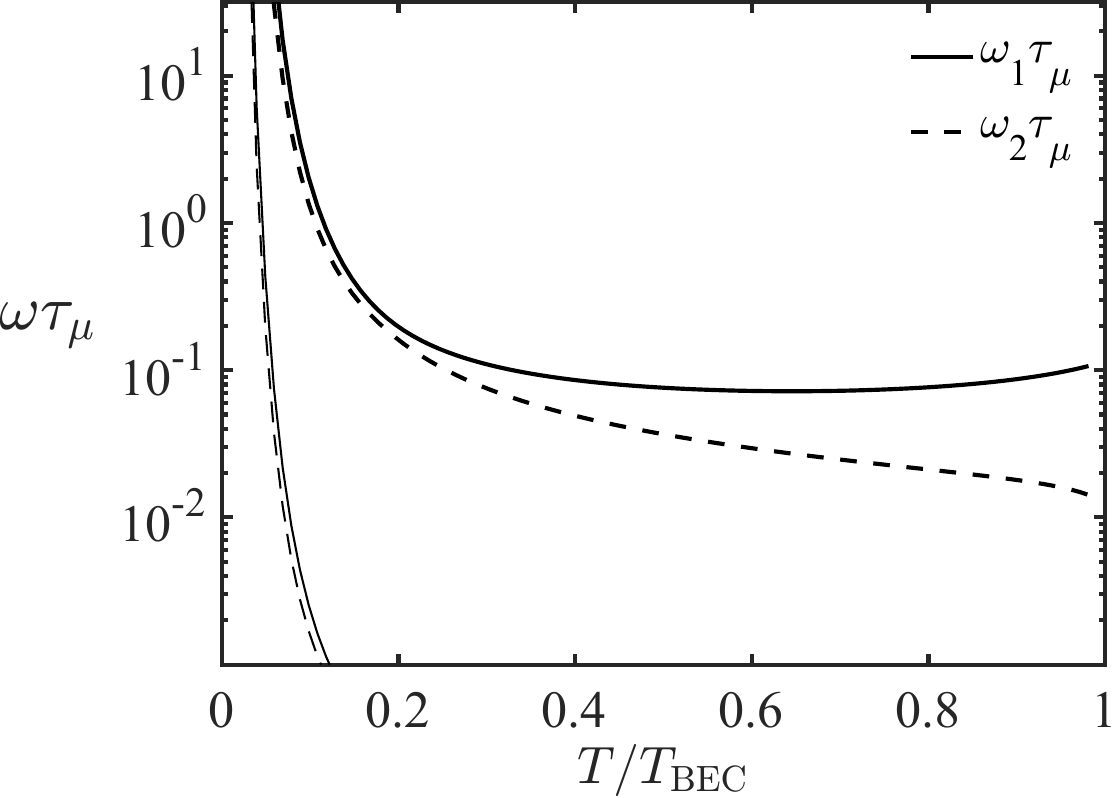}
    \caption{The dimensionless parameter $\omega\tau_{\mu}$ as a function of temperature.
    The value of $q\Lambda_{\rm BEC}$ is $4\times10^{-5}$ (thin line) and $q\Lambda_{\rm BEC}=0.0319$ (thick line). The eigenfrequency of the collective mode
    $\omega_{1,2}$ is approximately obtained from Eq.~\eqref{extendedzgn} and the relaxation time $\tau_{\mu}$ is given by Eq.~\eqref{chemicalpotentialrelaxationtime}.}
    \label{fig:omegatau}
\end{figure}

Figure~\ref{fig:omegatau} shows $\omega_{1,2}\tau_{\mu}$, where $\omega_{1,2}$ is the eigenfrequency in the hydrodynamic limit evaluated by  Eq.~\eqref{extendedzgn} and the $\tau_{\mu}$ is the relaxation time defined by Eq.~\eqref{chemicalpotentialrelaxationtime}.
% \textcolor{red}{We approximate the eigenfrequencies $\omega_{1,2}$ by the ones obtained from the hydrodynamic limit since}, in the moment method for finite-temperature Bose gases, extracting collective modes from eigenfrequencies of the moment equations is quite a difficult task because there are many eigenmodes around the collective excitations, that are not coupled to the density fluctuations. 
% However, the comparison in Fig.~\ref{fig:chiwq2d3d}~(b) shows that the Landau two-fluid model in Eq.~\eqref{extendedzgn} can accurately predict the frequency of the collective modes throughout the crossover regime.
We note that the comparison in Fig.~\ref{fig:chiwq2d3d}~(b) shows that the eigenfrequencies predicted by the dissipationless Landau two-fluid model in Eq. (63) are located around the peak of those predicted by the moment method for all the wavenumber $q$. 
Thus, we shall make use of this observation and approximate the eigenfrequencies $\omega_{1,2}$ by the Landau two-fluid model Eq.~\eqref{extendedzgn}, even in the crossover regime.
Moreover, we choose the representative relaxation time as $\tau_\mu$. 
This  is because, although we are dealing with the generalized relaxation rates $\gamma^l_{\alpha,kk'}$, the moment ZNG equation reduces to the dissipationless Landau two-fluid equations with the limit $\omega\tau_{\mu}\ll1$ as we saw in Sec.~\ref{sect:two-fluid}.

The hydrodynamic region ($\omega\tau \ll 1$) can be achieved, except the region very close to the zero temperature, by choosing the wavenumber as $q\Lambda_{\rm BEC}=4\times10^{-5}$ as shown in Fig.~\ref{fig:omegatau} thin(-dashed)-line.
% At $q\Lambda_{\rm BEC}=4\times10^{-5}$, except the region close to the zero-temperature, the collective modes are the hydrodynamic modes because $\omega\tau \ll 1$ is satisfied (thin lines in Fig.~\ref{fig:chiwT_hydro}). 
As expected, $\omega_{1,2}\tau_\mu$ increases with decreasing temperature, indicating that the collective modes are in the collisionless regime. 

\begin{figure}[tbp]
    \centering
       \includegraphics[scale=.4]{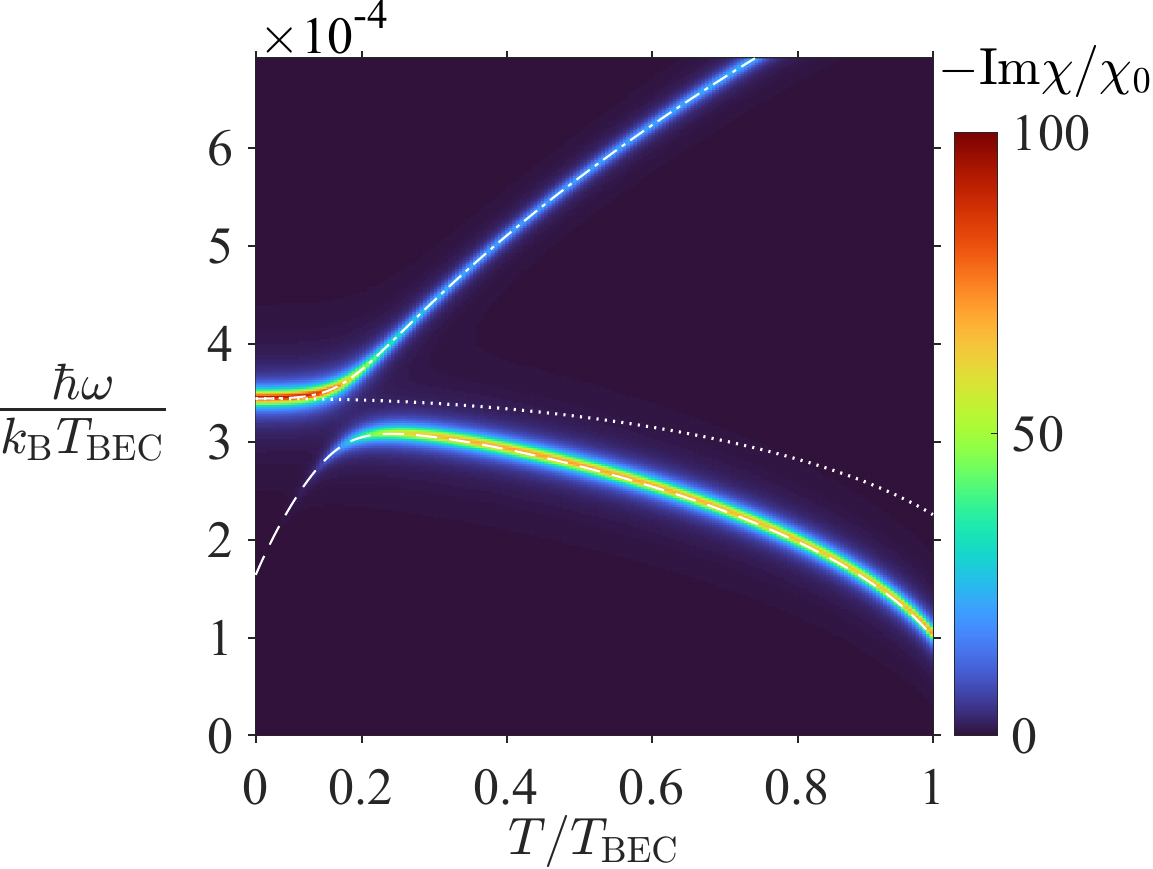} 
    \caption{The imaginary part of the density response function as the function of the frequency  $\omega$ and temperature $T$  at $q\Lambda_{\rm BEC}=4\times10^{-5}$. The rest of the parameter choice is the same as Fig.~\ref{fig:chiwq2d3d}.
    }
    \label{fig:chiwT_hydro}
\end{figure}

The corresponding density response function obtained by the moment ZNG equations is shown in Fig.~\ref{fig:chiwT_hydro}.
The response peak is located on the prediction given by Landau two-fluid equations at all the temperatures.
At high temperatures, the weight of the density response function is dominated by the second sound, although the first sound also emerges. 
Around $T\approx0.2T_{\rm BEC}$, the two modes are hybridized. 
At lower temperatures, the first sound smoothly crossovers to the collisionless Bogoliubov sound, and this branch dominates the density response function. 

By taking the wavenumber $q\Lambda_{\rm BEC}=0.0319$, the crossover is achieved at the higher temperature.
The dimensionless parameter $\omega\tau$ is plotted as thick lines in Fig.~\ref{fig:omegatau}.
While the first sound is relatively in the crossover regime $\omega\tau_\mu\sim1$ even at high temperatures, the second sound is almost in the hydrodynamic regime $\omega\tau<10^{-1}$ above $T\sim 0.2T_{\rm BEC}$.

\begin{figure}[tbp]
    \centering
    \includegraphics[scale=.4]{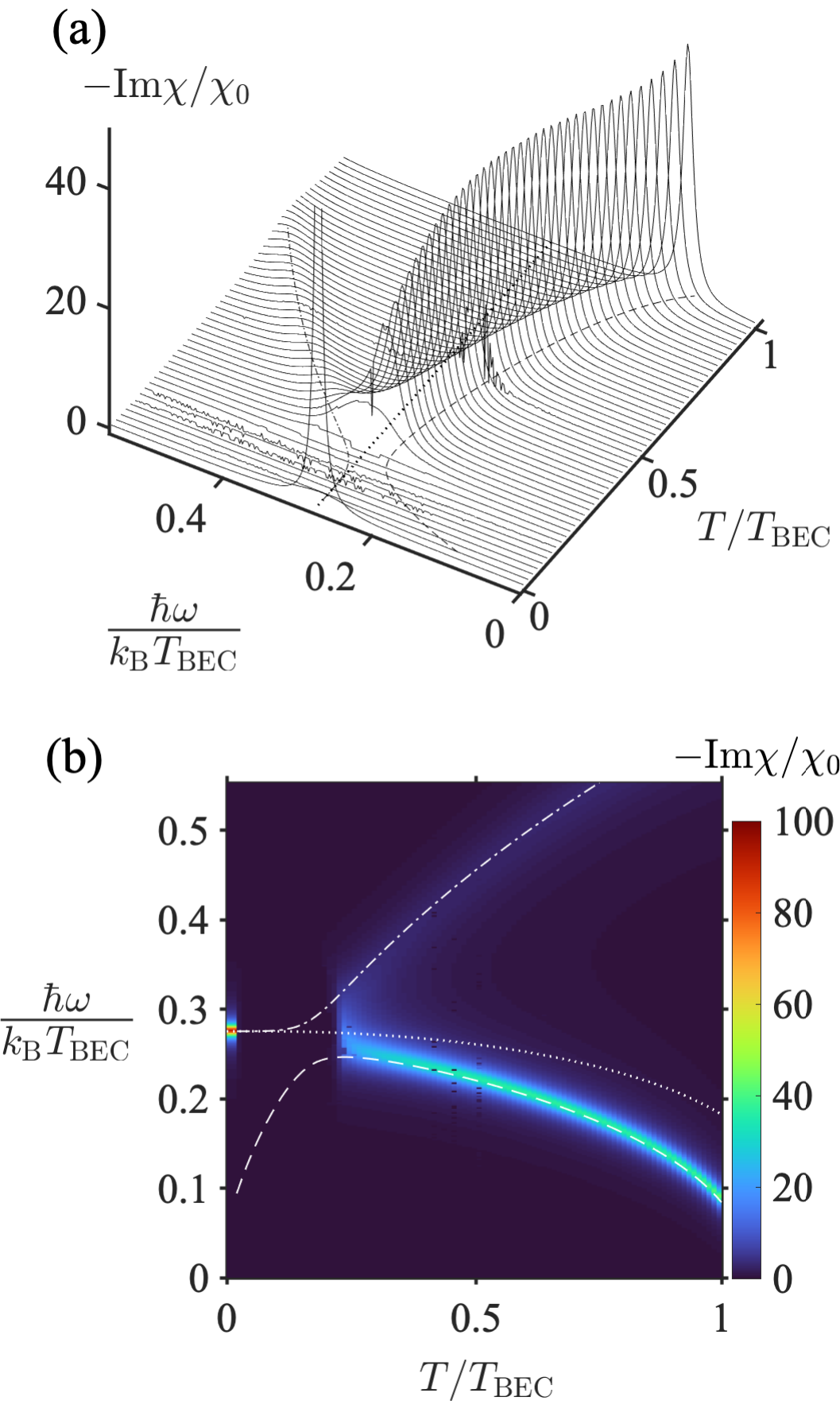}
    \caption{
    The imaginary part of the density response function as the function of the frequency  $\omega$ and temperature $T$ at $q\Lambda_{\rm BEC}=0.0319$. 
    The dotted line is the Bogoliubov sound frequency.
    The dot-dashed and dashed lines are the first and second sound frequencies obtained from Eq.~\eqref{extendedzgn}.
    } 
    \label{fig:chiwT}
\end{figure}

In Figs.~\ref{fig:chiwT} (a) and (b), we plot the temperature and frequency dependence of the imaginary part of the density response function obtained from the moment method. 
In the high-temperature regime, both the first and second sound peaks coincide with the eigenfrequencies in the hydrodynamic limit given by Eq.~\eqref{extendedzgn}. 
As noted in the discussion on Fig.~\ref{fig:chiwq2d3d}, even in the crossover regime with a small noncondensate fraction, the density response function is peaked at the first-sound frequency predicted by the Landau two-fluid equations. 
As can be seen in Figs.~\ref{fig:chiwT} (a) and (b), this observation holds for all the temperatures above the hybridization temperature. 

Around the temperature $T\approx 0.2T_{\rm BEC}$, the two modes are hybridized.
As seen in Fig.~\ref{fig:chiwT} (a), the response peaks from the first and second sounds merge around this temperature and lose weight at lower temperatures.
 Figures~\ref{fig:omegatau} and~\ref{fig:chiwT} (b) show that, at this temperature, both first and second sound modes are in the crossover regime, where no collective excitation is visible. 
This crossover region is relatively broad compared with the case shown in Fig.~\ref{fig:chiwT_hydro}. 
At sufficiently low temperatures, the Bogoliubov sound appears as expected. 

Here, we comment on the relation between the second-sound density response weight and the superfluid density.
In the two-dimensional Bose gas, the observation of the second sound is used to detect the superfluid density and draw the critical temperature of the Berezinskii--Kosterlitz--Thouless transition, whereas
in the case of the dilute Bose gas in three dimensions, the superfluid density is almost the same as a condensate density~\cite{yellowbook}.
Although the second sound (and the first sound around the hybridization temperature) loses the response weight due to the coupling to the thermal incoherent modes in the crossover regime, this does not imply the absence of the superfluid density.

In the present paper, we could not find evidence of the relaxational mode predicted by the ZNG equation~\cite{canada}, which is expected to appear in the dynamical structure factor at $\omega=0$, analogous to the classical thermal diffusion mode \cite{hu}. 
This is because one needs a short relaxation time $\tau_{\mu}$ to observe the relaxational mode, which is not achievable in current uniform ultracold atomic BECs.

\begin{figure}[tbp]
    \centering
    \includegraphics[scale=.4]{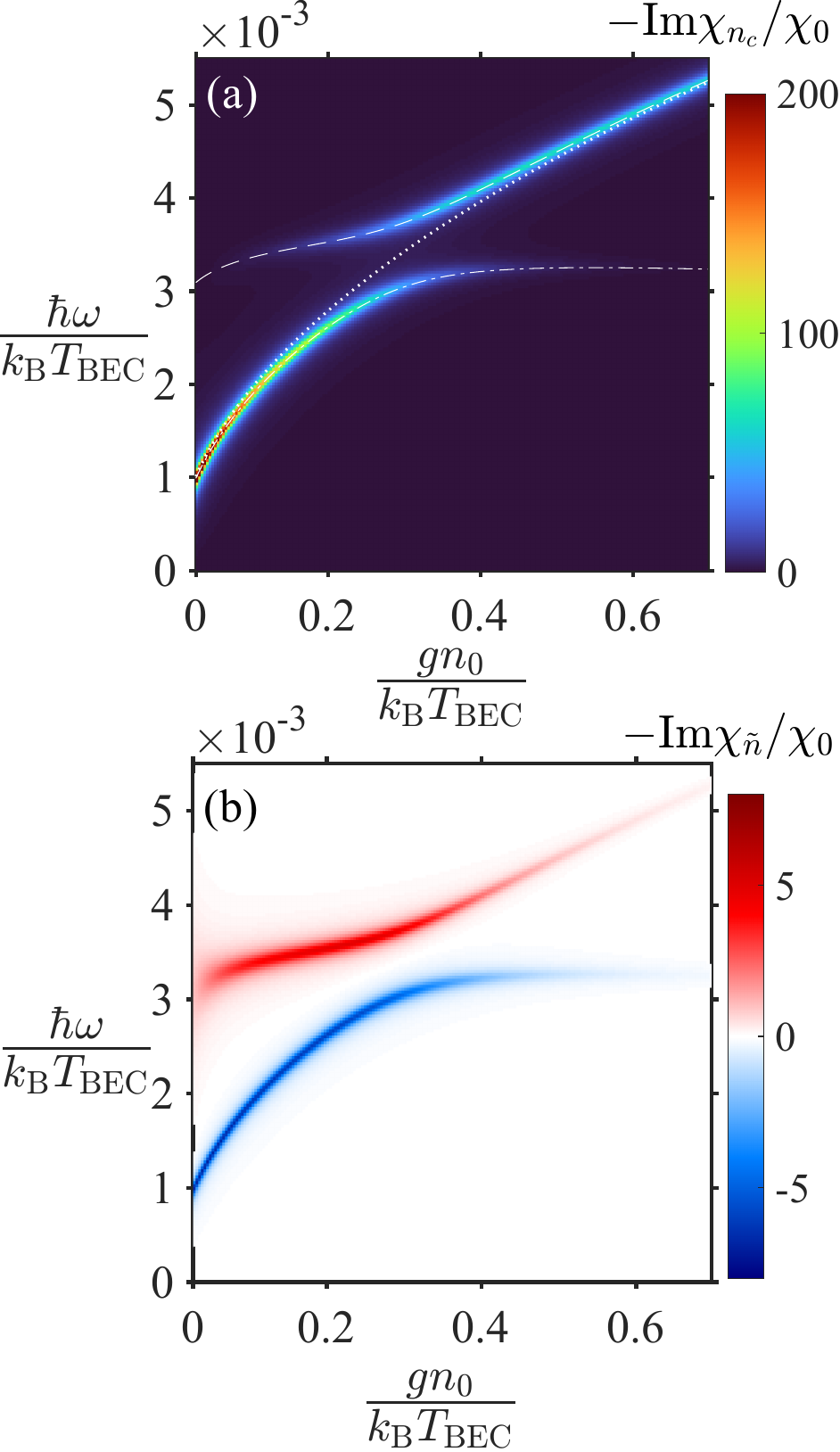} 
    \caption{Imaginary part of the (a) condensate and (b) noncondensate response function.
    The temperature is $T=0.2T_{\rm BEC}$ and wavenumber is $q\Lambda_{\rm BEC}=0.004\Lambda_{\rm BEC}$.
    The dotted line shows the frequency of the Bogoliubov sound.
    The dashed and dot-dashed lines are the first and second sound frequencies obtained by Eq.~\eqref{extendedzgn}.
    }
    \label{fig:chiwg}
\end{figure}

Figure \ref{fig:chiwg} shows the interaction-strength dependence of the density response function, where this parameter is controllable in the ultracold atomic gases through Feshbach resonance. 
Strictly speaking, the Hartree--Fock mean-field approximation used in the present paper requires  $gn_0\ll k_{\rm B}T_{\rm BEC}$. 
However, as long as the system is at a much lower temperature than the critical temperature, we can address the crossover between the collisionless and hydrodynamic regimes by changing the interaction parameters. 
Figure \ref{fig:chiwg} shows the imaginary part of (a) condensate and  (b)  noncondensate response functions defined in Eq.~\eqref{condensateresponse} and \eqref{noncondensateresponse}, respectively.
The hybridization occurs around $gn_0\sim 0.3k_{\rm B}T_{\rm BEC}$, which is consistent with the case in Fig.~\ref{fig:chiwT}. 
One can see that the noncondensate response function $- {\rm Im} \chi_{\tilde n}$ takes negative values in the case of the second sound, which reflects the out-of-phase oscillations. 
In the relatively large interaction strength region, the condensate response function keeps a significant contribution to the first sound, as shown in Fig.~\ref{fig:chiwg} (a), whereas the contribution of the noncondensate density response function to the first sound is reduced as shown in Fig.~\ref{fig:chiwg} (b).

%% --------------------------------------------

% -----------------------[ CONCLUSION ]------------------------- %	
\section{conclusion}\label{sect:conclusion} 

We developed the framework of the moment method applicable to collective sound modes in the finite-temperature BECs by using the coupled equations for the condensate and noncondensate.
The relaxation-time approximation in the moment method is discussed in detail in this paper. 
In the collision-dominated hydrodynamic limit, the truncated moment equations are shown to be equivalent to the Landau two-fluid equations.
This paper complements the study of the two-fluid sound in the hydrodynamic limit with \cite{mappelink,Arahata_2013} and without \cite{ZNG_original} the harmonic trap.

This paper provides an alternative approach to the standard Chapman--Enskog theory~\cite{chapmannenskog}.
Unlike the Chapman--Enskog theory, the moment method has the advantage of dealing with the collisionless limit by simply considering higher moments. 
We numerically solved the moment equations to investigate the crossover from the hydrodynamic to collisionless regimes. 
Unlike the prediction of the dissipative Landau two-fluid theory, the moment method uncovered that the response weight of the second sound is significantly reduced due to incoherent excitations. 
As a result, in the crossover region, the collective mode that gives the dominant weight switches between the first and second sound modes.
It would be possible to observe the behavior of the $\bm q$ dependence of the second sound using the experimental scheme recently developed by Hilker et.al~\cite{Hilker_2022} by simply varying the excitation wavevector, which was fixed to be comparable to the (cylindrical) box-trap size in Ref.~\cite{Hilker_2022}.
Comparing the results from the moment method with those from the Landau two-fluid equations, we found that the dimensionless parameter $\omega_{1,2}\tau_\mu$ can be used as an indicator of the hydrodynamic and collisionless regimes, where $\omega_{1,2}$ are first and second sound frequencies obtained from the Landau two-fluid equations and $\tau_{\mu}$ is the relaxation time associated with equilibration between the condensate and noncondensate.

We also investigated the hybridization of the first and second sounds in the crossover regime. 
Interestingly, the observed response function is qualitatively different from Landau two-fluid hydrodynamic theory. 
Finally, we found that the relaxational mode predicted by the ZNG two-fluid equations \cite{canada} is not visible in the dynamic structure factor because of small $\tau_{\mu}$.
In the typical experimental setup, harmonic traps are used to confine atomic clouds.
This makes the local condensation density larger than in homogeneous systems and makes $\tau_{\mu}$ smaller~\cite{Nikuni_2001}, making the observation of relaxation modes much more difficult  (see Eq.~\eqref{chemicalpotentialrelaxationtime}). 
% observation of the relaxational mode quite difficult since the local condensate density is larger than the homogeneous system, resulting in the smaller $\tau_{\mu}$ (see Eq.(62)).

The moment method systematically developed in this paper can potentially be extended to the various systems described by the Boltzmann equation with or without BECs.
In particular, studying the crossover of the collective excitation of the dipolar gas systems, where sound velocity depends on the alignment of the dipoles~\cite{Wang_2021,Wang_2022,Wang_2023}, would be an important application for future work.
%% ------ Answer to question 4 -------
Moreover, studying the crossover between the hydrodynamic and collisionless regime of two-dimensional Bose gas, where Berezinskii--Kosterlitz--Thouless transition plays a crucial role~\cite{Ville_2018,Ota_2018,Christodoulou_2021}, is an important application of the moment method.
However, since the ZNG scheme assumes the existence of the Bose-condensate order parameter, it cannot be applied directly to the description of such a two-dimensional system.
A suitable theoretical framework would be the classical field theory~\cite{davis_01,Blakie_2005,Bisset_2009}.
The extension of the ZNG scheme to describe the classical coherent field including the collisional process with the incoherent field is important future work and the moment method would be a powerful tool to describe the crossover between hydrodynamic and collisionless regimes.

\begin{acknowledgments}
This work has been supported by the Okinawa Institute
of Science and Technology Graduate University and used
the computing resources of the Scientific Computing and
Data Analysis section of the Research Support Division at Okinawa Institute of Science and Technology Graduate University.
H. H. would like to thank Kazuma Ohi, Yoji Asano, and Juan Polo Gomez for the valuable discussions.
S.W. was supported by Japan Society for the Promotion of Science KAKENHI Grant No. JP18K03499.
\end{acknowledgments}

\appendix

\section{Expansion of the linearized collision term}
\label{appendix:expansion_of_collision_term}
In the present paper, we expand the linearized collision term by the spherical harmonics.
In this appendix, we give the derivation of Eq.~\eqref{collisionterm_expandbyYlm}.

One can rewrite the linearized collision integral in terms of integral kernels as follows~\cite{Gust_13}:
\begin{align}
    \mathcal{L}_{22}[\nu(1)]=-\int \frac{d\bm p_2}{(2\pi\hbar)^3}K_{22}(1,2)\nu(2),
    \label{appendix:L_22kernel}
\end{align}
where the kernel $K_{22}(1,2)$ is defined as 
\begin{align}
	K_{22}(1,2) \equiv&
	(2\pi\hbar)^3\delta(\bm p_1-\bm p_2)M_{22}(1)\notag\\
        &+Q_{22}(1,2)-2S_{22}(1,2),
\end{align}
in conjunction with
\begin{align}
	&Q_{22}(1,2)\equiv
	\int \frac{d\bm p_3}{(2\pi\hbar)^3}\int d\bm p_4W_{22}(1,2,3,4),
	\\
	&S_{22}(1,3)\equiv
	\int \frac{d\bm p_2}{(2\pi\hbar)^3}\int d\bm p_4W_{22}(1,2,3,4),
	\\
	&M_{22}(1)\equiv
	\int \frac{d\bm p_2}{(2\pi\hbar)^3}\int \frac{d\bm p_3}{(2\pi\hbar)^3}\int d\bm p_4W_{22}(1,2,3,4).
\end{align}
Similarly, one can express $\mathcal{L}_{12}$ as 
\begin{align}
	\mathcal{L}_{12}[\nu(1)]=
	-\int\frac{d\bm p_2}{(2\pi\hbar)^3}K_{12}(1,2)\nu(2),
 \label{appendix:L_12kernel}
\end{align}
where the kernel $K_{12}(1,2)$ is given by
\begin{align}
	K_{12}(1,2)\equiv&
	Q_{12}(1,2)-2S_{12}(1,2),\\
	Q_{12}(1,2)=&\int d\bm p_3\int d\bm p_4W_{12}(1,2,3,4),\\
	S_{12}(1,3)=&\int d\bm p_2\int d\bm p_4W_{12}(1,2,3,4).
\end{align}

Taking into account the spherical symmetry in the momentum space, we find that the kernel $M_{22}(1)$ merely depends on $p_1$, whereas $K_{\alpha}(1,2)$ depend on $p_1$, $p_2$, and $\hat{\bm p}_1\cdot\hat{\bm p}_2=\cos\theta$.
Therefore, we expand $K_{\alpha}(1,2)$ by the Legendre polynomial:
\begin{align}
	&K_\alpha(1,2)=\sum_{l}\frac{2l+1}{4\pi}K^l_\alpha(p_1,p_2)P_l(\cos\theta),
 \label{appendix:expand_kernel}
\end{align}
where the expansion coefficient is given by $K^l_\alpha(p_1,p_2)=2\pi\int^1_{-1}d(\cos\theta)K_\alpha(1,2)P_l(\cos\theta)$.
Making use of the spherical harmonic addition theorem, one can rewrite Eq.~\eqref{appendix:expand_kernel} as
\begin{align}
	K_\alpha(1,2)=\sum_{l}K^l_\alpha(p_1,p_2)\sum^l_{m=-l}
	Y^{*}_{l,m}(\hat{\bm p}_2)Y_{l,m}(\hat{\bm p}_1),
 \label{appendix:expand_kernel_sphericalharmonics}
\end{align}
where the expansion coefficient is given by
$K^l_\alpha(p_1,p_2)=\int d\hat{\bm p}_1\int d\hat{\bm p}_2Y^{*}_{l,0}(\hat{\bm p}_1)K_\alpha(1,2)Y_{l,0}(\hat{\bm p}_2)$.
Inserting Eq.~\eqref{appendix:expand_kernel_sphericalharmonics} and \eqref{expandbysphericalharmonics} into Eq.~\eqref{appendix:L_22kernel} and \eqref{appendix:L_12kernel}, we obtain
\begin{align}
	\mathcal{L}_\alpha(\bm p_1)
 =&\sum_l\sum^l_{m=-l}\sqrt{\frac{4\pi}{2l+1}}\mathcal{L}^{ml}_{\alpha}(p_1)Y_{l,m}(\hat{\bm p}_1),\\
 \mathcal{L}^{ml}_{\alpha}(p_1)=&-\int\frac{p^2_2dp_2}{(2\pi\hbar)^3}K^l_{\alpha}(p_1,p_2)\nu^m_l(p_2).
\end{align}
In this way, it is possible to expand the linearized collision term directly by the spherical harmonics.

\section{relaxation-time approximation}\label{appendix:relaxationtimeapprox} 

In this appendix, we shall show the derivation Eq.~\eqref{momenteq_l0k0}-\eqref{momenteq_l2} in detail. 
First, the polynomial expansion of the fluctuation $\nu_l$ leads to the following expression for the moment and collision integral:
\begin{align}
    \expval{p^{l+2k}\nu_l}
    =&\sum_{k'}C^l_{k'}W_{2l+2k+2k'},
    \label{appendix:momentintermofsoninepolynomial}\\
    J_{l}[p^{l+2k},\nu_{l}(p)]
    =&
    \sum_{\alpha}\sum_{k'}\gamma^l_{\alpha,kk'}W_{2l+2k+2k'}C^l_{k'}.
    \label{appendix:JinSoninePolynomial}
\end{align}
For simplicity, we specify the $(l,k)$ th order of moment equation as the $(l,k)$-moment equation. 

Let us consider the $(0,0)$-moment equation.
The collisional process of $C_{22}$ conserves the number of particle, which gives the relation $J_{0,22}[1,\nu]=	J_{0,22}[\nu,1]=0$ for any function $\nu$. 
Thus the collision term of the $(0,0)$-moment equation only involves $C_{12}$ collision rate $\gamma^0_{12,0k'}$, given by 
\begin{align}
J_{0}[1,\nu_0]
=\sum_{k'}\gamma^0_{12,0k'}W_{2k'}C^0_{k'}.
\label{appendix:l0k0collisionterm}
\end{align}

The moments $\expval{\nu_0}$ and $\expval{p^2\nu_0}$ are related to the fluctuation of the number of particles and energy in local equilibrium (see Appendix.~\ref{appendix:leq}).
When one develops the relaxation-time approximation, the contribution from such moments has to be included.
This can be done systematically as follows.
We expand the summation in Eq.~\eqref{appendix:l0k0collisionterm} up to $k'=2$, and for $k'>2$, we replace the relaxation rate $\gamma^0_{12,0k'}$ with $\gamma^0_{12,02}$. 
This is the relaxation-time approximation used in this paper. 
After this replacement, we obtain 
\begin{align}
	J_{0}[1,\nu_0]
	=&
 \gamma^0_{12,00}W_0C^0_{0}+\gamma^0_{12,01}W_2C^0_{1}\notag \\
 &+\gamma^0_{12,02}\pqty{\sum_{k'}W_{2k'}C^0_{k'}-W_0C^0_0-W_{2}C^0_{1}}\notag \\
	=&(\gamma^0_{12,00}-\gamma^0_{12,02})W_0C^0_{0}\notag\\
 &+(\gamma^0_{12,01}-\gamma^0_{12,02})W_2C^0_{1}+\gamma^0_{12,02}\expval{\nu_0},
	\label{appendix:Jl0k0}
\end{align}
where Eq.~\eqref{appendix:momentintermofsoninepolynomial} is used.
We use the local equilibrium solution (see Appendix.~\ref{appendix:leq})
\begin{align}
	\nu_0^{\rm leq}= C^0_{0}+C^0_{1}p^2, 
 \label{appendix:nuleql0k0}
\end{align}
to determine $C^0_0$ and  $C^0_1$. 
Using Eq. \eqref{appendix:nuleql0k0}, we have relations 
\begin{align}
	\expval{\nu_0}\approx&\expval{\nu^{\rm leq}_0} =C^0_0W_0+C^0_1W_2,\\
	\expval{p^2\nu_0}\approx&\expval{\nu^{\rm leq}_0} = C^0_0W_2+C^0_1W_4.
\end{align} 
We then obtain 
\begin{align}
	C^0_0=&\frac{1}{D}[W_4\expval{\nu_0}-W_2\expval{p^2\nu_0}],
 \label{appendix:l0k0mometleq_sol_C00}
 \\
	C^0_1=&\frac{1}{D}[W_0\expval{p^2\nu_0}-W_2\expval{\nu_0}],
 \label{appendix:l0k0mometleq_sol_C01}
\end{align}
where $D\equiv W_0W_4-W^2_2$. 
By inserting the above local equilibrium solution to Eq.~\eqref{appendix:Jl0k0}, one finds
\begin{align}
	J_{0}[1,\nu_0]=&\bqty{\gamma^0_{12,00}-\frac{W^2_2}{D}(\gamma^0_{12,01}-\gamma^0_{12,00})}\expval{\nu_0}\notag\\
 &+\frac{W_2W_0}{D}(\gamma^0_{12,01}-\gamma^0_{12,00})\expval{p^2\nu_0}.
 \label{appendix:l0k0collision_RTA}
\end{align}
Substituting this form of the collision integral to Eq.~\eqref{rhsofmomenteq}, we obtain the $(0,0)$-moment equation as in Eq.~\eqref{momenteq_l0k0}. 
As in Sec.~\ref{sect:linearzation} and Appendix \ref{appendix:two-fluid}, the above equation is exactly the same as the continuity equation for the noncondensate atoms Eq.~\eqref{landauzngtwofluidcontinuity}. This justifies our choice of the relaxation-time approximation in Eq.~\eqref{appendix:Jl0k0}.

The collision term of the $l=0$ and $k=1$ moment equation can be related to the energy conservation law, 
which is given by 
\begin{align}
	J_{0,12}\bqty{\frac{p^2}{2m},\nu_0}=-gn_{c0}J_{0,12}\bqty{1,\nu_0}
\end{align}
and  $J_{0,22}\bqty{p^2,\nu}=J_{0,22}\bqty{\nu,p^2}=0$ for any function $\nu$.
Following the same procedure as in the $(0,0)$-moment equation, we obtain the following relation 
\begin{align}
	J_{0}\bqty{p^2,\nu_0}
		=&\frac{W_2W_4}{D}\pqty{\gamma^0_{12,01}-\gamma^0_{12,11}}\expval{\nu_0}\notag\\
  &+\pqty{\frac{W_4W_0}{D}\gamma^0_{12,11}-\frac{W^2_2}{D}\gamma^0_{12,01}}\expval{p^2\nu_0}\label{appendix:l0k1collision_RTA},
\end{align}
where we used the relation between two relaxation times given by
\begin{align}
\gamma^0_{12,01}=&\frac{J_{0,12}[1,p^2]}{W_{2}}
	=-2mgn_{c}\frac{W_0}{W_{2}}\gamma^0_{12,00},\\
\gamma^0_{12,11}=&\frac{J_{0,12}[p^2,p^2]}{W_{4}}=-2mgn_{c0}\frac{W_2}{W_4}\gamma^0_{12,01}.
\end{align}
This leads us to find the Eq.~\eqref{momenteq_l0k1}.

The above analysis can be extended to the $(0,k)$-moment equation at $k \geq 2$. 
Using the relaxation-time approximation, where we replace $\gamma^0_{\alpha,kk'}$ with $\gamma^0_{\alpha,22}$ and using the local equilibrium solution in Eqs.~\eqref{appendix:l0k0mometleq_sol_C00} and \eqref{appendix:l0k0mometleq_sol_C01}, we find
\begin{align}
	J_0[p^{2k},\nu_0]=
		&\biggl[(\gamma^0_{12,0k}-\gamma^0_{12,22}-\gamma^0_{22,22})\frac{W_{2k}W_4}{D}\notag\\
  &-(\gamma^0_{12,1k}-\gamma^0_{12,22}-\gamma^0_{22,22})\frac{W_{2k+2}W_2}{D}\biggr]\expval{\nu_0}\notag \\
	&+\biggl[(\gamma^0_{12,1k}-\gamma^0_{12,22}-\gamma^0_{22,22})\frac{W_{2k+2}W_0}{D}\notag\\
	&-(\gamma^0_{12,0k}-\gamma^0_{12,22}-\gamma^0_{22,22})\frac{W_{2k}W_2}{D}
 \biggr]
	\expval{p^2\nu_0}\notag \\
	&+(\gamma^0_{12,22}+\gamma^0_{22,22})\expval{p^{2k}\nu_0}. 
%	\label{momenteq_l0k}
\end{align}
We intentionally keep the $k$-depenendence in the collision term $\gamma^0_{12,0k}$ and $\gamma^0_{12,1k}$ in order to reproduce the local equilibrium solution.

In the case of the $(1,0)$-moment equation, 
the momentum conservation law reduces the collision integral as $J_{1,\alpha}[p,\nu]=J_{1,\alpha}[\nu,p]=0$ for both $C_{12}$ and $C_{22}$ collision processes.
Therefore, this mode only contributes to the local equilibrium solution of the system, and the moment equation is given in Eq.~\eqref{momenteq_l1k0}.

In the case of the $(1,k)$-moment equation for $k\neq0$, we take the relaxation-time approximation, where we replace $\gamma^1_{\alpha,kk'}$ for $k'\neq0$ with $\gamma^1_{\alpha,11}$. Then, we find
\begin{align}
	J_1[p^{1+2k},\nu_1]
	=&-(\gamma^1_{12,11}+\gamma^1_{22,11})W_{2+2k}C^1_{0}\notag\\
 &+(\gamma^1_{12,11}+\gamma^1_{22,11})\expval{p^{1+2k}\nu_1}. 
 \label{appendix:J_1afterSonineexpansion}
\end{align}
In order to express $C^1_0$ in terms of the moment, we use the local equilibrium solution $	\nu_1\approx\nu^{\rm leq}_1= C^1_0p$, which provides $\expval{p\nu_1}=W_2C^1_0$. Substituting the result into Eq.~\eqref{appendix:J_1afterSonineexpansion}, one finds
\begin{align}
J_1[p^{1+2k},\nu_1]
	=&-(\gamma^1_{12,11}+\gamma^1_{22,11})\frac{W_{2+2k}}{W_2}\expval{p\nu_1}\notag\\
 &+(\gamma^1_{12,11}+\gamma^1_{22,11})\expval{p^{1+2k}\nu_1}, 
\end{align}
which provides the moment equation in Eq.~\eqref{momenteq_l1k}.

Finally, in the $(l,k)$-moment equation for $l\geq2$,
we replace the  collision rate $\gamma^l_{\alpha,k'k}$ for $l\geq2$ with $\gamma^2_{\alpha,00}$. 
We thus find only the diagonal term 
\begin{align}
	J_l[p^{l+2k},\nu_l]
 =&
	\sum_{\alpha}\gamma^2_{\alpha,00}\sum_{k'}W_{2l+2k+2k'}C^l_{k'}\notag\\
 =& ( \gamma^2_{12,00}+\gamma^2_{22,00} ) \expval{p^{l+2k}\nu_l}, 
\end{align}
which provides Eq.~\eqref{momenteq_l2}.

In general, the diffusive local equilibrium solution in Eq.~\eqref{appendix:leqsolfromdistributionfunc} is defined as the solution of  $d\expval{p^{l+2k}\nu^{\rm leq}_l}_{\rm coll}/dt=0$. 
The collision term with the relaxation-time approximation developed above satisfies this condition for all $l$ and $k$.
Moreover, we explicitly showed that the moment equation with the relaxation-time approximation becomes equivalent to the Landau two-fluid equations in the hydrodynamic limit in Sec.~\ref{sect:two-fluid}.

%% ~~~~~~~~~~~~~~~~~~~~~~~~~~~~~~~~~~~~~~~~~~~

%% --------------------------------------------
\section{Hydrodynamic approximation of moment equation: Landau Two-fluid equation}\label{appendix:two-fluid}

In this appendix, we present the details of the reduction of the moment equation to the Landau two-fluid equations in the hydrodynamic regime.
Our starting point is the moment equation with hydrodynamic variables in Eq.~\eqref{landauzngtwofluidcontinuity}-\eqref{momentl0k1coursegrainedvar}.
Due to the hierarchical structure of the moment equation,
Eq.~\eqref{landauzngtwofluidcontinuity}-\eqref{momentl0k1coursegrainedvar} are not closed, but coupled to the higher order of moments through $\expval{p^{2}\nu_{2}}$ or $\expval{p^{3}\nu_{1}}$. 

In the following, we show that considering the hydrodynamic approximation, one can obtain the closed set of two-fluid equations.
For notational conciseness, we omit the sub(super)-script zero.

To find the approximate expression of $\expval{p^{2}\nu_{2}}$, we consider the $l\geq2$ moment equation.
 Defining the total relaxation time 
 \begin{align}
     \frac{1}{\tau_\eta}=\gamma^2_{12,00}+\gamma^2_{22,00},
     \label{appendix:taueta}
 \end{align}
  in Eq.~\eqref{momenteq_l2},
 we find
\begin{align}
	\omega\tau_\eta\expval{p^{l+2k}\nu_l}=&\tau_\eta \biggl ( \frac{q}{m}\frac{l}{2l-1}\expval{p^{l+2k+1}\nu_{l-1}}\notag\\
	&+\frac{q}{m}\frac{l+1}{2l+3}\expval{p^{l+2k+1}\nu_{l+1}} \biggr )\notag\\
	&-i\expval{p^{l+2k}\nu_l}.
\end{align}

The lowest approximation gives the local equilibrium solution.
That is, in the hydrodynamic limit, where all the relaxation times are extremely short, we can let $\tau_\eta$ be zero, which leads to $\expval{p^{l+2k}\nu_l}=0$. 
This lowest approximation leads to the two-fluid equations without shear viscosity (see Eq.~\eqref{appendix:p2nu2momentbyviscosityandvelocity}). 
Most importantly, this approximation also gives the physical meaning of the truncation of the moment equation. 
When one truncates the moment equation, the higher moments will be set to zero such that the resulting hierarchy is closed.
This effectively means that we take the local equilibrium solution for the moments higher than the cutoff.

We consider the first order approximation with respect to the small $\omega\tau_\eta$, which gives the deviation from the local equilibrium solution, given by 
\begin{align}
	i\expval{p^{l+2k}\nu_l}=&
	\tau_\eta \biggl ( \frac{q}{m}\frac{l}{2l-1}\expval{p^{l+2k+1}\nu_{l-1}}\notag\\
	&+\frac{q}{m}\frac{l+1}{2l+3}\expval{p^{l+2k+1}\nu_{l+1}} \biggr ).
 \label{appendix:generalexpressiondeparturefromleql2}
\end{align}
The first and second terms on the right-hand side are coupled with the $(l-1,k+1)$- and $(l+1,k)$-moment equation, respectively. 
Subsequently using Eq.~\eqref{appendix:generalexpressiondeparturefromleql2}, we see that the right-hand side has the factor of $\tau^n_{\eta}$, where $n(\geq2)$ depends on the truncation of the moment equation, and we reproduce the linearized Boltzmann equation with $n\to\infty$.
Therefore, in the two-fluid hydrodynamic regime, the effect of the $l>2$ moments would become exponentially small and decouple from the lower order of moments.

However, the first-order correction to the moment $\expval{p^2\nu_2}$ from Eq.~\eqref{appendix:generalexpressiondeparturefromleql2} is important since this is directly coupled to the fluctuation of the noncondensate velocity field $\delta v_n$. Using Eq.~\eqref{appendix:generalexpressiondeparturefromleql2}, the correction for  $\expval{p^2\nu_2}$ is given by
\begin{align}
	i\expval{p^{2}\nu_2}\approx&
 \tau_\eta\frac{2q}{3m}\expval{p^{3}\nu_{1}},
 \label{appendix:deviationfromleqp2nu2}
\end{align}
where we ignored $\expval{p^3\nu_3}$, since this term is of the order of $\tau^2_\eta$.
Defining the total relaxation rate
\begin{align}
    \frac{1}{\tau_{\kappa}}=\gamma^1_{12,11}+\gamma^1_{22,11},
    \label{appendix:taukappa}
\end{align}
and taking $\omega\tau_\kappa = 0$ in Eq.~\eqref{momenteq_l1k}, we find for the $(1,1)$-moment equation
\begin{align}
    i\expval{p^{3}\nu_1}=&i\frac{W_{4}}{W_2}\expval{p\nu_1}
	+\tau_\kappa\frac{q\beta}{m}W_{4}(2g\delta n_c+2g\expval{\nu_0}+U_{\rm ext})\notag \\
 &+\tau_{\kappa}\frac{q}{m}\expval{p^{4}\nu_{0}}+\tau_{\kappa}\frac{2q}{5m}\expval{p^{4}\nu_{2}}.
 \label{appendix:deviationfromleql1k}
\end{align}
This is the correction of $\expval{p^{3}\nu_1}$ to the first order in $\tau_{\kappa}$, which couples to $\expval{p^4\nu_0}$. 
Substituting this into Eq.~\eqref{appendix:deviationfromleqp2nu2} and letting $O(\tau_{\kappa}\tau_{\eta})$ be zero, we find the first order correction of $\expval{p^2\nu_2}$, as in Eq.~\eqref{p2nu2momentbyviscosityandvelocity}, proportional to the noncondensate velocity field, given by 
\begin{align}
	\expval{p^{2}\nu_2}=
	-i\tau_{\eta} \frac{q}{m}\frac{2}{3}\frac{W_{4}}{W_2}\expval{p\nu_1}
 =-10im\eta q\delta v_n,
 \label{appendix:p2nu2momentbyviscosityandvelocity}
\end{align}
where we defined the shear viscosity
\begin{align}
    \eta=\frac{\tau_{\eta}\beta_0W_4}{15m^2}.\label{appendix:viscosity}
\end{align}
Different from the transport coefficients obtained by Chapman--Enskog approach \cite{chapmannenskog}, the relaxation time related to the shear viscosity is given by the relaxation rate of (2,0)-moment equation, $\gamma^2_{12,00}$ and $\gamma^2_{22,00}$. Although the analytical form of the relaxation time related to the shear viscosity in Eq.~\eqref{appendix:viscosity} and the one derived in the Ref. \cite{chapmannenskog} are different, we numerically confirmed that both formulas have similar temperature dependence.

To estimate the first-order correction of $\expval{p^3\nu_1}$, we shall use the $(1,0)$-moment equation in Eq.~\eqref{momenteq_l1k0}, given by 
\begin{align}
	&\beta\frac{q}{m}W_{2}[2g\delta n_c+2g\expval{\nu_0}+U_{\rm ext}]\notag\\
 &=\omega\expval{p\nu_1}-\frac{q}{m}\expval{p^{2}\nu_{0}}-\frac{2q}{5m}\expval{p^{2}\nu_{2}}.
\end{align} 
Substituting this into Eq.~\eqref{appendix:deviationfromleql1k} with the hydrodynamic limit approximation $\omega\tau_\kappa=0$, we find the relation 
\begin{align}
	\expval{p^{3}\nu_1}=&\frac{W_{4}}{W_2}\expval{p\nu_1}-i\tau_\kappa\biggl[\frac{q}{m}\expval{p^{4}\nu_{0}}\notag\\
	&
 -\frac{W_{4}}{W_2}\pqty{\frac{q}{m}\expval{p^{2}\nu_{0}}+\frac{2q}{5m}\expval{p^{2}\nu_{2}}}\biggr],
 \label{appendix:p3nu1momentequationafterhydrolimit}
\end{align}
where we neglected
the term proportional to  $\expval{p^{4}\nu_{2}}$, since this is an order of $\tau_\eta$, as one can see from Eq.~\eqref{appendix:generalexpressiondeparturefromleql2}. 
The moments $\expval{p\nu_1}$, $\expval{p^2\nu_0}$, and $\expval{p^2\nu_2}$ are given by Eqs.~\eqref{appendix:vel_p1nu1}, \eqref{appendix:P_p2nu0} and \eqref{appendix:p2nu2momentbyviscosityandvelocity}, respectively. 
The lowest approximation $\tau_{\kappa}\to0$ gives the local equilibrium solution that leads to the two-fluid equations without thermal conductivity (see Eq.~\eqref{appendix:p3nu1momentbykappa}). 

To determine the moment $\expval{p^3\nu_1}$ to the first order in $\tau_\kappa$, we need the local equilibrium solution of the moments $\expval{p^4\nu_0}$. 
Defining $\gamma^0_{12,00}+\gamma^0_{22,00}=1/\tau_{\rm tot}$, and taking $\omega\tau_{\rm tot}\to0$ with Thomas--Fermi approximation in  Eq.~\eqref{momenteq_l0k}, we have
\begin{align}
	i\expval{p^{4}\nu_0}=&\tau_{\rm tot}\frac{q}{3m}\frac{W_{6}}{W_2}\expval{p\nu_1}-i\beta g\tau_{\rm tot}\gamma^0_{12,02}W_{4}\delta n_c\notag \\
		&-i \biggl [ (\tau_{\rm tot}\gamma^0_{12,02}-1)\frac{W_{4}W_4}{D}\notag\\
  &-(\tau_{\rm tot}\gamma^0_{12,12}-1)\frac{W_{6}W_2}{D} \biggr ] \expval{\nu_0}\notag \\
	&-i \biggl [ ( \tau_{\rm tot}\gamma^0_{12,12}-1 ) \frac{W_{6}W_0}{D}\notag\\
	&- ( \tau_{\rm tot}\gamma^0_{12,02}-1 )\frac{W_{4}W_2}{D} \biggr ]
	\expval{p^2\nu_0}. 
 \label{appendix:p4nu0momentequationafterhydrolimit}
\end{align}
Taking the hydrodynamic limit $\omega\tau_{12} = 0$ in the $(0,0)$-moment equation in Eq.~\eqref{landauzngtwofluidcontinuity}, we have the equation 
\begin{align}
	i\beta g\delta n_c=\frac{q\tau_{12}}{3mn_{c}}\expval{p\nu_{1}}
	-i\frac{\sigma_{2}}{\tilde n}\expval{\nu_0}-i\frac{\sigma_{1}\beta}{\tilde n}\frac{\expval{p^2\nu_0}}{3m}.
\end{align}
Substituting this result into Eq.~\eqref{appendix:p4nu0momentequationafterhydrolimit}, we have 
\begin{align}
	i\expval{p^{4}\nu_0}
	=&-i\pqty{\frac{W_{6}W_2}{D}-\frac{W^2_4}{D}}\expval{\nu_0}\notag\\
 &+i\bqty{\frac{W_0}{D}\pqty{W_{6}-\frac{W_4^2}{W_2}}+\frac{W_4}{W_2}}\expval{p^2\nu_0}\notag\\
 &+ {\mathcal O}(\tau_{\rm tot}), 
\end{align}
where we used $\gamma^0_{12,12}=-2mgn_{c}(W_4/W_{6})\gamma^0_{12,02}$ and $\expval{p\nu_1} = {\mathcal O}(\tau_{\rm tot})$.
Substituting this into Eq.~\eqref{appendix:p3nu1momentequationafterhydrolimit}, we obtain Eq.~\eqref{p3nu1momentbykappa}, given by 
\begin{align}
	\expval{p^{3}\nu_1}=&
 15m^2\tilde P\delta v_n+i\frac{6q\sigma_{4}m^2T\kappa}{\tilde n}\delta\tilde n\notag\\
	&-i\frac{6q\sigma_{3}m^2T\kappa}{\tilde P}\delta\tilde P,
 \label{appendix:p3nu1momentbykappa}
\end{align}
where we have replaced the moments with the hydrodynamic variables in Eq.~\eqref{momentsbydensityvelocitypressurefluctuation} and introduced the thermal conductivity given by 
\begin{align}
	\kappa= \tau_\kappa\frac{k_B\beta^2}{12m^4}\pqty{W_6-\frac{W^2_4}{W_2}}.
\end{align}

%% ~~~~~~~~~~~~~~~~~~~~~~~~~~~~~~~~~~~~~~~~~~~~~~
\section{Collision integral in moment equation}\label{appendix:collisionintegral}

In this appendix, we provide an explicit expression for the collision integral defined by Eq.~\eqref{def_moment_relaxation_rate}.
Let us introduce a dimensionless momentum variable $\bm p=\sqrt{2mk_BT_0}\bm\xi$, and rewrite the collision integral as
\begin{align}
J_{l,22}[p^{l+2k},p^{l+2k'}]
=
(2l+1)C(T)\mathcal{I}_{22},
\label{appendix:Jl22_byCandI22}
\end{align}
where the dimensionless integral $\mathcal{I}_{22}$ and the coefficient $C(T)$ are given by 
\begin{align}
	\mathcal{I}_{22} \equiv &\int d\bm\xi_1\int d\bm\xi_2\int d\bm\xi_3\int d\bm\xi_4\notag\\
 &\times
 \delta(\bm\xi_1+\bm\xi_2-\bm\xi_3-\bm\xi_4)\delta(\xi^2_1+\xi^2_2-\xi^2_3-\xi^2_4)\notag \\
	&\times f^0(1)f^0(2)[1+f^0(3)][1+f^0(4)]\xi^{l+2k}_1P_l(\cos\theta_1)\notag \\
&\times[\xi^{l+2k'}_1P_l(\cos\theta_1)+\xi^{l+2k'}_2P_l(\cos\theta_2)\notag\\
&-\xi^{l+2k'}_3P_l(\cos\theta_3)-\xi^{l+2k'}_4P_l(\cos\theta_4)],\label{appendix:I_22}
\\
C(T) \equiv &\frac{(2mk_BT)^{9/2+l+k+k'}}{(2\pi\hbar)^9k_BT}\frac{4\pi g^2}{\hbar}.
\label{appendix:coef_of_I22}
\end{align}
where $f^0=\qty[z^{-1}_0\exp(\beta_0\frac{p^2}{2m})-1]^{-1}$ with the fugacity $z_0=e^{\beta_0 (\mu_0-2gn_0 )}$ is the equilibrium Bose distribution function.
Following Refs. \cite{nikunitransportcoefC22,chapmannenskog}, we shall introduce a variable transformation given by
\begin{align}
	&\bm\xi_1=(\bm\xi_0+\bm\xi')/\sqrt{2},
 \label{appendix:variable_transfomation_x1}
 \\
	&\bm\xi_2=(\bm\xi_0-\bm\xi')/\sqrt{2},\\
	&\bm\xi_3=(\bm\xi'_0+\bm\xi'')/\sqrt{2},\\
	&\bm\xi_4=(\bm\xi'_0-\bm\xi'')/\sqrt{2}.
 \label{appendix:variable_transfomation_x4}
\end{align}
With this variable transformation,  the delta function appearing in Eq.~\eqref{appendix:I_22} can be written as $\delta(\bm\xi_1+\bm\xi_2-\bm\xi_3-\bm\xi_4)=\delta[\sqrt{2}(\bm\xi_0-\bm\xi'_0)]$ and 
$\delta(\xi^2_1+\xi^2_2-\xi^2_3-\xi^2_4)=\delta(\xi'^2-\xi''^2)$. 

We introduce the polar coordinate to $\bm\xi_0$, given by 
\begin{align}
	\bm\xi_0=\xi_0 (\sin\theta_0\cos\phi_0,\sin\theta_0\sin\phi_0,\cos\theta_0).
 \label{appendix:polar_xi0}
\end{align} 
Given all the variables appearing in Eq.~\eqref{appendix:variable_transfomation_x1}-\eqref{appendix:variable_transfomation_x4} with polar coordinates similar to Eq.~\eqref{appendix:polar_xi0}, the integral in Eq.~\eqref{appendix:I_22} can be reduced to a numerical calculation friendly form with eight variables after the integration of the delta function.
One can reduce the integral further by defining the following local coordinate: 
\begin{figure}
    \centering
     \includegraphics[scale=.5]{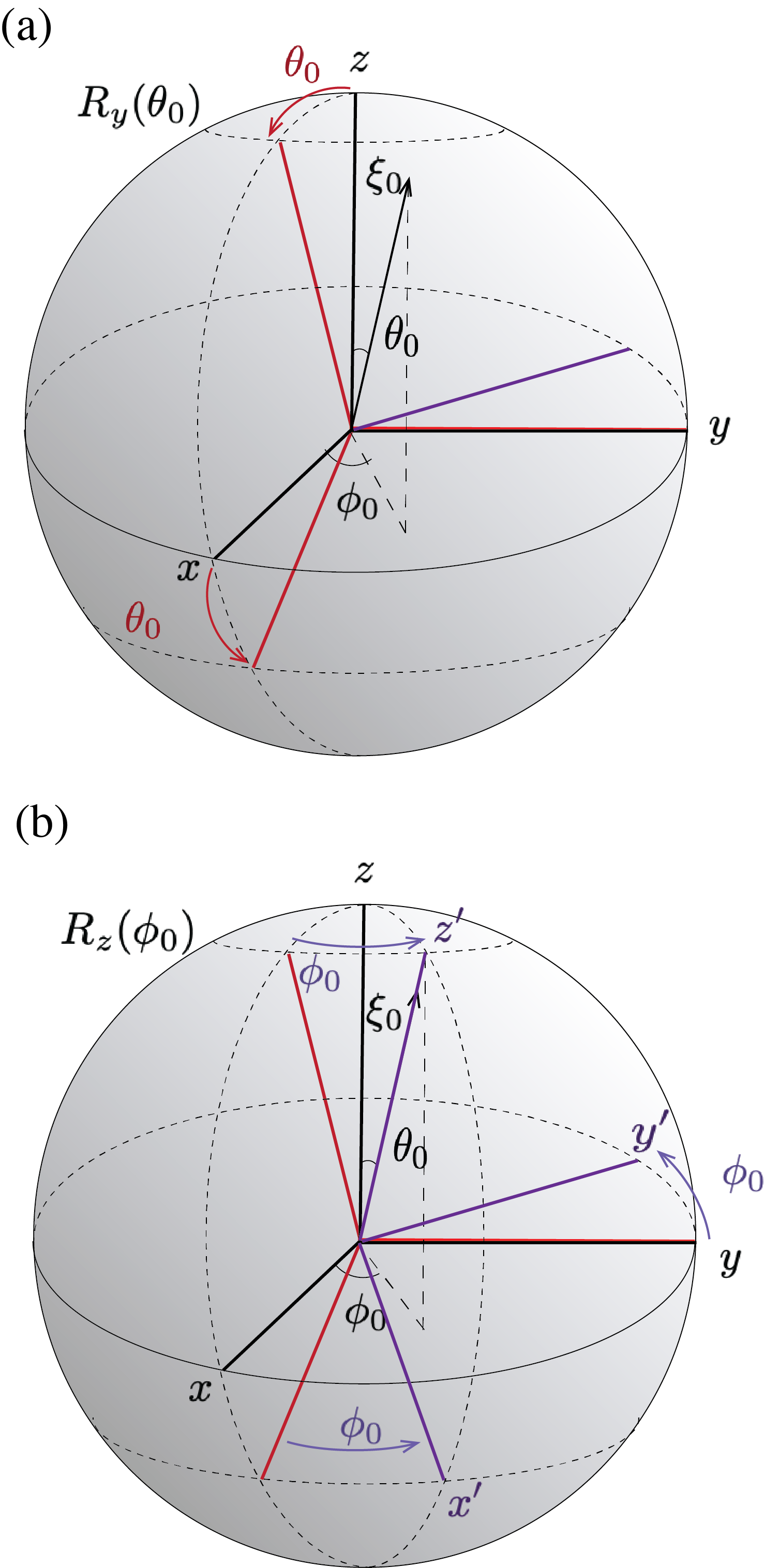}
    \caption{The variable transformation by the matrix $R$ given by Eq.~\eqref{appendix:rotationmat}. 
    }
    \label{fig:rotation}
\end{figure}
We define the local coordinate system $(x',y',z')$ for $\bm\xi'$ such that  the $z'$-axis coincides with $\bm\xi_0$, then one can express $\bm\xi'$ in terms of the polar coordinates of $\bm\xi_0$ as 
\begin{align}
	\xi'_x=&\xi'_{x'}\cos\theta_0\cos\phi_0-\xi'_{y'}\sin\phi_0\notag\\
 &+\xi'_{z'}\sin\theta_0\cos\phi_0,
 \label{appendix:xi_prime_x_bylocalcoord}
 \\
	\xi'_y=&\xi'_{x'}\cos\theta_0\sin\phi_0-\xi'_{y'}\cos\phi_0\notag\\
 &+\xi'_{z'}\sin\theta_0\sin\phi_0,
 \label{appendix:xi_prime_y_bylocalcoord}
 \\
	\xi'_z=&-\xi'_{x'}\sin\theta_0+\xi'_{z'}\cos\theta_0. 
 \label{appendix:xi_prime_z_bylocalcoord}
\end{align}
In terms of the rotation matrix, this can be written as
\begin{align}
    \mqty(
    \xi'_x\\
    \xi'_y\\
    \xi'_z
    )
    =R_z(\phi_0)R_y(\theta_0)\mqty(
    \xi'_{x'}\\
    \xi'_{y'}\\
    \xi'_{z'}
    ),
\end{align}
where $R_y$ and $R_z$ are  the rotation matrices around $y$ and $z$ axis, respectively.
The two successive rotations $R \equiv R_z(\phi_0)R_y(\theta_0)$ can be written as
\begin{align}
    R=\mqty(
    \cos\phi_0\cos\theta_0 & -\sin\phi_0 & \cos\phi_0\sin\theta_0\\
    \sin\phi_0\cos\theta_0 & \cos\phi_0 & \sin\phi_0\sin\theta_0\\
    -\sin\theta_0 & 0 & \cos\theta_0
    ).
    \label{appendix:rotationmat}
\end{align}
The variable transformation by $R$ is schematically shown in Fig.~\ref{fig:rotation}.
Expressing $(\xi'_{x'},\xi'_{y'},\xi'_{z'})$ in terms of the polar coordinate defined in the local coordinate system $(x',y',z')$, one also has
\begin{align}
	\xi'_{x'}=&\xi'\sin\theta'\cos\phi',\\
	\xi'_{y'}=&\xi'\sin\theta'\sin\phi',\\
	\xi'_{z'}=&\xi'\cos\theta'.
\end{align}
Substituting the above expressions into Eq.~\eqref{appendix:xi_prime_z_bylocalcoord}, one has $	\xi'_z=-\xi'\sin\theta_0\sin\theta'\cos\phi'+\xi'\cos\theta_0\cos\theta'.$
Using $\xi_{1z}=(\xi_{0z}+\xi'_z)/\sqrt{2}$, we obtain
\begin{align}
	\xi_{1z}
	=\frac{\xi_0\cos\theta_0+\xi'(\cos\theta_0\cos\theta'-\sin\theta_0\sin\theta'\cos\phi')}{\sqrt{2}}. 
 \label{appendix:xi1z}
\end{align} 

On the other hand, the polar coordinate of $\xi_1$ defined in the global coordinate system $(x,y,z)$ can be written as $\xi_{1z}=\xi_1\cos\theta_1$. Thus, inserting this into Eq.~\eqref{appendix:xi1z}, we find the angle $\cos\theta_1$ as follows
\begin{align}
	\cos\theta_1=\frac{\xi_0\cos\theta_0+\xi'(\cos\theta_0\cos\theta'-\sin\theta_0\sin\theta'\cos\phi')}{\sqrt{2}\xi_1}.
\end{align}
The expression of $\xi_1$ in terms of $\bm\xi_0$ and $\bm\xi'$ can be written as 
\begin{align}
	\xi^2_1=\frac{1}{2}(\xi^2_0+\xi'^2+2\bm\xi_0\cdot\bm\xi'),
\end{align}
where owing to the definition of the local coordinate system $(x',y',z')$, we have 
\begin{align}
    \bm\xi_0\cdot\bm\xi'
    =\mqty(
    \xi_{0x} &\xi_{0y} & \xi_{0z} 
    )
    R
    \mqty(
    \xi'_{x'} \\
    \xi'_{y'} \\
    \xi'_{z'} 
    )
    =\xi_0\xi'\cos\theta',
\end{align}
where Eqs.~\eqref{appendix:polar_xi0},~\eqref{appendix:xi_prime_x_bylocalcoord}-\eqref{appendix:xi_prime_z_bylocalcoord}, and~\eqref{appendix:rotationmat} are used. 
In this way, one can express $\cos\theta_1$ in terms of the new coordinate variables $\xi_0,\xi',\theta_0,\theta'$, and $\phi'$.
In the same manner one can obtain $\xi^2_{2,3,4}$, which can be summarized as 
\begin{align}
	&\xi^2_1=\frac{1}{2}(\xi^2_0+\xi'^2+2\xi_0\xi'\cos\theta'),\\
	&\xi^2_2=\frac{1}{2}(\xi^2_0+\xi'^2-2\xi_0\xi'\cos\theta'),\\
	&\xi^2_3=\frac{1}{2}(\xi^2_0+\xi'^2+2\xi_0\xi'\cos\theta''),\\
	&\xi^2_4=\frac{1}{2}(\xi^2_0+\xi'^2-2\xi_0\xi'\cos\theta''),
\end{align}
and angle components $y_i=\cos\theta_i$ given by 
\begin{align}
	&y_1=\frac{\xi_0\cos\theta_0+\xi'(\cos\theta_0\cos\theta'-\sin\theta_0\sin\theta'\cos\phi')}{\sqrt{2}\xi_1},\\
	&y_2=\frac{\xi_0\cos\theta_0-\xi'(\cos\theta_0\cos\theta'-\sin\theta_0\sin\theta'\cos\phi')}{\sqrt{2}\xi_2},\\
	&y_3=\frac{\xi_0\cos\theta_0+\xi'(\cos\theta_0\cos\theta''-\sin\theta_0\sin\theta''\cos\phi'')}{\sqrt{2}\xi_3},\\
	&y_4=\frac{\xi_0\cos\theta_0-\xi'(\cos\theta_0\cos\theta''-\sin\theta_0\sin\theta''\cos\phi'')}{\sqrt{2}\xi_4}.
\end{align}

Introducing another variable transformation $\xi_0=\sqrt{2\eta}\cos\psi$ and $\xi'=\sqrt{2\eta}\sin\psi$, where $\eta>0$ and $\psi\in[0,\pi/2]$, the function $F(\xi_0,\xi',y',y'')=f^0(1)f^0(2)[1+f^0(3)][1+f^0(4)]$ can be rewritten as 
\begin{align}
	&F(\eta,\psi,y,y'')\notag \\
	&=\frac{1}{4}\frac{1}{\cosh[\eta-\beta_0\mu_0+2\beta_0gn_0]-\cosh(\eta y'\sin2\psi)}\notag\\
	&\times\frac{1}{\cosh[\eta-\beta_0\mu_0+2\beta_0gn_0]-\cosh(\eta y''\sin2\psi)}. 
\end{align}
Using the above variable transformations, the collision integral can be reduced to 
\begin{align}
	\mathcal{I}_{22}
=&2\pi\int^\infty_{0} d\eta\int^{\pi/2}_0d\psi\int^1_{-1}dy_0\int^1_{-1} dy'\int^1_{-1} dy''\notag\\
&\times \int^{2\pi}_0d\phi'\int^{2\pi}_0d\phi''
\eta^{l+k+k'+5/2}\cos^2\psi\sin^3\psi \notag\\
&\times F(\eta,\psi,y',y'';z)\zeta^{l+2k}_1P_l(y_1)\notag \\
&\times\biggl[\zeta^{l+2k'}_1P_l(y_1)+\zeta^{l+2k'}_2P_l(y_2)\notag\\
&-\zeta^{l+2k'}_3P_l(y_3)-\zeta^{l+2k'}_4P_l(y_4)
\biggr],
\label{appendix:final_expression_I22}
\end{align}
where we defined $\xi^2_i=\eta\zeta^2_i$. 
For $l=0$, we have $P_{l=0} (y_i) = 1$, which is irrelevant to $y_i$, and one can integrate out the variables $y_0,\phi'$ and $\phi''$ analytically. 
Then, we recover the expression for the collision integral previously developed for the Bose gas~\cite{nikunitransportcoefC22,chapmannenskog,yellowbook}.
% and for the normal Fermi gas~\cite{watabeosawanikuni}.

Similarly, one can reduce the collision integral $J_{l,12}[p^{l+2k},p^{l+2k'}]$ to 
\begin{align}
	J_{l,12}&[p^{l+2k},p^{l+2k'}]=(2l+1)
n_{c0}\Lambda^3_0C(T)\mathcal{I}_{12},
\end{align}
where the temperature-dependent coefficient $C(T)$ is given by Eq.~\eqref{appendix:coef_of_I22}, and the collision integral  $\mathcal{I}_{12}$ is given by
\begin{align}
	\mathcal{I}_{12}\equiv&
	\pi^{3/2}\int d\bm\xi_1
\int d\bm\xi_2\int d\bm\xi_3\int d\bm\xi_4
\delta(\bm\xi_2-\bm\xi_3-\bm\xi_4)
\notag \\
&\times
\delta(\xi^2_2-\xi^2_3-\xi^2_4+\beta_0\mu_{c0}-2\beta_0gn_0)\notag\\
&\times[\delta(\bm\xi_1-\bm\xi_2)-\delta(\bm\xi_1-\bm\xi_3)-\delta(\bm\xi_1-\bm\xi_4)]\notag \\
&\times [1+f^0(2)]f^0(3)f^0(4)\notag\\
&\times\xi^{l+2k}_1P_l(y_1)
[\xi^{l+2k'}_2P_l(y_2)-2\xi^{l+2k'}_3P_l(y_3)] .
\end{align}
Introducing the variable transformation
\begin{align}
	\bm\xi_2=\frac{1}{2}(\bm\xi_0+\bm\xi_r),
	\;\;\;\;\;
	\bm\xi_3=\frac{1}{2}(\bm\xi_0-\bm\xi_r),
\end{align}
and following the method developed in the calculation of the integral $\mathcal{I}_{22}$, 
we find
\begin{align}
	\xi^2_2&=\frac{1}{4}(\xi^2_0+2\xi_0\xi_r\cos\theta_r+\xi^2_r),\\
	\xi^2_3&=\frac{1}{4}(\xi^2_0-2\xi_0\xi_r\cos\theta\cos\theta_r+\xi^2_r),\\
	y_2&=\frac{1}{2\xi_2}[\xi_0y_0+\xi_r(y_0y_r-\sin\theta_0\sin\theta_r\cos\phi_r)], \\
	y_3&=\frac{1}{2\xi_3}[\xi_0y_0-\xi_r(y_0y_r-\sin\theta_0\sin\theta_r\cos\phi_r)]. 
\end{align}
Performing all the integrals involving the delta functions, we find
\begin{align}
	\mathcal{I}_{12}
=&\pi^{3/2}\frac{\pi}{4}
\int d\xi_r\int^{1}_{-1} dy\int^{1}_{-1} dy_0\int^{2\pi}_0d\phi_r\notag\\
&\times\xi^2_r\sqrt{\xi^2_r+2\beta_0gn_{c0}}F(\xi_r,y,z_0) \notag \\
&\times[\xi^{l+2k}_1P_l(y_0)-\xi^{l+2k}_2P_l(y_2)-\xi^{l+2k}_3P_l(y_3)]
\notag \\
&\times[\xi^{l+2k'}_1P_l(y_0)-\xi^{l+2k'}_2P_l(y_2)-\xi^{l+2k'}_3P_l(y_3)],
\label{appendix:final_expression_I12}
\end{align}
where the function $F=[1+f^0(1)]f^0(2)f^0(3)$ is given by
\begin{align}
	 F(\xi_r,y,z_0)=\frac{z_0e^{-\xi^2_1}}{(1-z_0e^{-\xi^2_1})(1-z_0e^{-\xi^2_2})(1-z_0e^{-\xi^2_3})},
\end{align}
where $\xi_i$ and $y_i$ can be given by 
\begin{align} 
	\xi^2_1=&\xi^2_0=\xi^2_r+2\beta_0gn_{c0},\\
	\xi^2_2=&(\xi^2_0+2\xi_0\xi_ry_r+\xi^2_r)/4\notag\\
	=&\frac{1}{2}(\xi^2_r+\beta_0gn_{c0}+\xi_ry_r\sqrt{\xi^2_r+2\beta_0gn_{c0}}),\\
	\xi^2_3=&(\xi^2_0-2\xi_0\xi_ry_r+\xi^2_r)/4\notag\\
	=&\frac{1}{2}(\xi^2_r+\beta_0gn_{c0}-\xi_ry_r\sqrt{\xi^2_r+2\beta_0gn_{c0}}),\\
	y_2=&\frac{y_0\sqrt{\xi^2_r+2\beta_0gn_{c0}}+\xi_r(y_0y_r-\sin\theta_0\sin\theta_r\cos\phi_r)}{2\xi_2},\\
	y_3=&\frac{y_0\sqrt{\xi^2_r+2\beta_0gn_{c0}}-\xi_r(y_0y_r-\sin\theta_0\sin\theta_r\cos\phi_r)}{2\xi_3}.
\end{align}

Figure~\ref{fig:relax_rates} shows the collision rate in Eq.~\eqref{def_moment_relaxation_rate} associated with the hierarchy of the moment equations.
For reference, we also plot the inverse relaxation time $1/\tau_{\alpha}$ for $\alpha = \{12, 22\}$ associated with the $C_{\alpha}$ collision process, given by 
\begin{align}
    \frac{1}{\tau_{12}}=&\frac{4\pi g^2}{\hbar}
    \int \frac{d\bm p_1}{(2\pi\hbar)^3}
    \int \frac{d\bm p_2}{(2\pi\hbar)^3}
    \int d\bm p_3
    \notag \\
    &\times\delta(\bm p_1-\bm p_2-\bm p_3)
    \delta(\mu_{c0}+\tilde\varepsilon_1-\tilde\varepsilon_2-\tilde\varepsilon_3)\notag\\
    &\times(1+f(1))f(2)f(3)
    ,\label{appendix:tau12}
    \\
    \frac{1}{\tau_{22}}=&\frac{4\pi g^2}{\tilde n_0\hbar}
    \int \frac{d\bm p_1}{(2\pi\hbar)^3}
    \int \frac{d\bm p_2}{(2\pi\hbar)^3}
    \int\frac{d\bm p_3}{(2\pi\hbar)^3}\int d\bm p_4\notag\\
    &\times\delta(\bm p_1+\bm p_2-\bm p_3-\bm p_4)\notag\\
    &\times\delta(\tilde\varepsilon(1)+\tilde\varepsilon(2)-\tilde\varepsilon(3)-\tilde\varepsilon(4))\notag\\
    &\times
    f^0(1)f^0(2)[1+f^0(3)][1+f^0(4)].
\end{align}
By using the explicit expression in Eq.~\eqref{appendix:final_expression_I12}, we can find the following formulas between two collision rates:
\begin{align}
    \gamma^0_{12,00}=&\frac{n_{c0}}{W_0\tau_{12}},
    \label{appendix:gamma000_tau}
    \\
    \gamma^0_{12,01}=&-\frac{2\beta_0gW_0n_{c0}}{3\tilde n_0}\gamma^0_{12,00},\\
    \gamma^0_{12,11}=&\frac{(2mgn_c)^2W_0}{W_4}\gamma^0_{12,00},\\
    \gamma^0_{12,12}=&-\frac{2mgn_{c0}W_4}{W_6}\gamma^0_{12,02}.
    \label{appendix:gamma012_gamma002}
\end{align}

\section{\label{appendix:tau_kappaeta}Relaxation time $\tau_{\kappa,\eta}$}

In Sec.\ref{sect:two-fluid}, truncating the moment equation, we obtained the Landau two-fluid equation, including dissipation from the thermal conductivity $\kappa$ and $\eta$.
As we mentioned in the main text, the associated relaxation time $\tau_{\kappa}$ and $\tau_{\eta}$ slightly differ from the ones obtained by employing the Chapman--Enskog approach~\cite{chapmannenskog,yellowbook}.
Here, we discuss this difference in detail.

The obtained thermal conductivity is associated with the characteristic time scale $\tau_{\kappa}$, given by
\begin{align}
    \frac{1}{\tau_{\kappa}}=\frac{1}{\tau^{\kappa}_{22}}+\frac{1}{\tau^{\kappa}_{12}},
\end{align}
where $\tau^{\kappa}_{12}$ and $\tau^{\kappa}_{22}$ originate from $C_{12}$ and $C_{22}$ collision processes.
To highlight the difference, we shall discuss the difference in $\tau^\kappa_{22}$.

To compare the relaxation time obtained by both methods, it is convenient to rewrite the collision integral Eq.~\eqref{def_moment_relaxation_rate} in terms of the Bose--Einstein function $g_{n}(z)$.
To this end, we use the formula
\begin{align}
    W_n=\frac{2h^n}{\pi^{1/2+n/2}\Lambda^{3+n}_0}\Gamma\pqty{\frac{n+3}{2}}g_{(n+1)/2}(z_0).
    \label{appendix:Wn_by_gn}
\end{align}
Inserting this formula into Eq.~\eqref{def_moment_relaxation_rate} along with Eq.~\eqref{appendix:Jl22_byCandI22}, we obtain
\begin{align}
    \frac{1}{\tau^\kappa_{22}}
    \equiv&
    \gamma^1_{22,11}\notag\\
    =&\frac{2}{15}\frac{m^3(k_BT_0)^2g^2}{\pi^{13/2}\hbar^7}\frac{3\mathcal{I}_{22}[\xi^3\cos\theta,\xi^3\cos\theta]}{\tfrac{7}{2}g_{7/2}(z_0)},
    \label{appendix:moment_Ikappa22}
\end{align}
where the detailed expression of the collision integral $\mathcal{I}_{22}$ is given by Eq.~\eqref{appendix:I_22} for $l=k=k'=1$.

Employing the Chapman--Enskog method, Nikuni and Griffin obtained the Landau two-fluid equations, including the transport coefficients $\kappa$ and $\eta$ \cite{chapmannenskog}.
The expression for $1/\tau^\kappa_{22}$ in this case is given by~\cite{yellowbook}
\begin{align}
    \frac{1}{\tau^\kappa_{22}}&=\frac{2}{15}\frac{m^3g^2(k_BT_0)^{2}}{\pi^{13/2}\hbar^{7}}\frac{I^\kappa_{22}(z_0)}{\frac{7}{2}g_{7/2}(z_0)\mathcal{D}},
    \label{appendix:ZNG_Ikappa22_numerator}
    \\
    \mathcal{D}&=g_{3/2}(z_0)-\frac{5g^2_{5/2}(z_0)}{7g_{7/2}(z_0)},
    \label{appendix:ZNG_Ikappa22_denominator}
\end{align}
where $I^\kappa_{22}$ is the collision integral associated with $1/\tau^\kappa_{22}$.
Defining the nondimensional linearized collisional operator
\begin{align}
    \mathcal{L}_{22}[\nu]=\frac{m^3(k_BT_0)^2}{2\pi^5\hbar^7}g^2\mathcal{L}'_{22}[\nu],
\end{align}
the collision integral $I^\kappa_{22}$ can be written as follows
\begin{align}
    I^\kappa_{22}
    =&
    -\int d\bm\xi_1\xi^2_1\bm
    \xi_1\cdot\mathcal{L}'_{22}[\xi^2_1\bm \xi_1]\notag\\
    =&-3\int d\bm\xi_1\xi^2_1
    \xi^z_1\mathcal{L}'_{22}[\xi^2_1 \xi^z_1],
\end{align}
where we used the spherical symmetry of the collision integral.
Introducing the spherical coordinate, we find
\begin{align}
    I^\kappa_{22}
    =&-3\int d\bm\xi_1\xi^3_1
    \cos\theta_1\mathcal{L}'_{22}[\xi^3_1\cos\theta_1]\notag\\
    =&3\mathcal{I}_{22}[\xi^3\cos\theta,\xi^3\cos\theta].
    \label{appendix:I_kappa22_by_mclI22}
\end{align}
Comparing Eq.~\eqref{appendix:moment_Ikappa22} and \eqref{appendix:ZNG_Ikappa22_numerator} along with \eqref{appendix:ZNG_Ikappa22_denominator} and \eqref{appendix:I_kappa22_by_mclI22},
we see that the relaxation time obtained by the two methods differs only from the factor in the denominator $\mathcal{D}$.

However, this is not the case for $\tau_{\eta}\equiv(1/\tau^\eta_{22}+1/\tau^\eta_{12})^{-1}$.
In this case, the associated collision integral from Chapman--Enskog theory for the $C_{22}$ collision process is given by 
\begin{align}
    I^\eta_{22}=\sum_{\mu,\nu=x,y,z}\int d\bm \xi\xi_\mu\xi_\nu\mathcal{L}'_{22}[\xi_\mu\xi_\nu]
    \equiv
    \sum_{\mu,\nu=x,y,z}I^{\mu\nu,\mu\nu}_{\eta,22},
\end{align}
whereas the collision integral from the moment method is given by $l=2,k=k'=0$ in Eq.~\eqref{appendix:I_22}.
Since the collision integral Eq.~\eqref{appendix:I_22} only depends on the polar angle $\theta$, it can only reproduce the diagonal element $I^{\mu\mu,\mu\mu}_{\eta,22}$.
The cross term, such as $I^{xy,xy}_{\eta,22}$, involves the azimuthal angle.

In this appendix, we highlighted the difference in the expression of $\tau_{\kappa}$ and  $\tau_{\eta}$ obtained by the moment method and the Chapman--Enskog approach.
Nonetheless, we emphasize that the resultant transport coefficients $\kappa$ and $\eta$ show similar temperature dependence~\cite{chapmannenskog,yellowbook}; thus, the resultant eigenfrequencies coincide almost perfectly.

\section{Local equilibrium solution}\label{appendix:leq}
In the appendix \ref{appendix:relaxationtimeapprox}, we developed the relaxation-time approximation by considering the deviation from the local equilibrium solution. 
This solution is also used to associate the moments with the physical quantities to derive the Landau two-fluid equations.
Here, we present the derivation of the local equilibrium solution.
The distribution function  $f^{(0)}$ is the local equilibrium solution if $f^{(0)}$ satisfies the following condition:
\begin{align}
    C_{12}[f^{(0)}]+C_{22}[f^{(0)}]=0.
\end{align}
The solution satisfying this condition is given by the local-equilibrium Bose distribution function
\begin{align}
    &f^{(0)}(\bm p,\bm r,t)\notag\\
    &=\frac{1}{
    \exp{\beta[(\bm p-m\bm v_n)^2/2m+U-\tilde\mu]} - 1
    },
\end{align}
where $\beta,\bm v_n,U,$ and $\tilde\mu$ all depend on the position $\bm r$ and time $t$.
Let us consider the deviation from the equilibrium distribution function $f^0$ in the uniform system \cite{yellowbook,watabeosawanikuni}:
\begin{align}
    f^{(0)}-f^0
    =f^0(1+f^0)\biggl[&\beta^2_0\pqty{\frac{p^2}{2m}+2gn_0-\tilde\mu_0}\delta\theta\notag\\
    &+\beta_0\delta\tilde{\mu}-2\beta_0g\delta n+\beta_0\bm p\cdot\bm v_n\biggr].
\end{align}

Comparing with the definition of the fluctuation of the distribution function $f^{(0)}-f^0=\nu f^0(1+f^0)$, we find that the fluctuation $\nu$ around the local equilibrium is given by
\begin{align}
    \nu^{\rm leq}(\bm p)
    =&\beta^2_0\pqty{\frac{p^2}{2m}+2gn_0-\tilde\mu_0}\delta\theta\notag\\
    &+\beta_0\delta\tilde{\mu}-2\beta_0g\delta n+\beta_0\bm p\cdot\delta\bm v_n . 
    \label{appendix:leqsolfromdistributionfunc}
\end{align}
In terms of the polynomial expansion Eqs.~\eqref{expandbysphericalharmonics} and ~\eqref{expandfluctuationbySonine}, one has a relation given by 
\begin{align}
    \nu(\bm p)=\sum_l\sum_kC^l_kp^{l+2k}Y_l(\hat{\bm p}).
    \label{appendix:sonineandsphericalharmonicsexpansion}
\end{align}
Therefore, one can write the local equilibrium solution as 
\begin{align}
    \nu^{\rm leq}(\bm p)=a+\bm b\cdot\bm p+cp^2,
    \label{appendix:leqsolbyexpansion}
\end{align}
where $a=C^0_0, \bm b\cdot\bm p=C^1_0pY_1(\hat{\bm p})$, and $c=C^0_1$. 

In a normal gas, we can determine the coefficients $C^0_0,\bm C^1_0$, and $C^0_1$ by using the conservation lows of the collision integrals \cite{watabeosawanikuni,asano}. 
% The starting point of the moment method, in this case, is to consider relaxation-time approximation with respect to the deviation from $\nu^{\rm leq}$.
Comparing Eq.~\eqref{appendix:leqsolbyexpansion} with Eq.~\eqref{appendix:leqsolfromdistributionfunc}, one can find the coefficients $a,{\bm b}$ and $c$ in terms of the fluctuations from equilibrium, given by 
\begin{align}
    a=&\beta gn_c\delta\theta-\beta g\delta n_c+\beta\delta\tilde\mu-2\beta_0g\delta n,\label{appendix:coef_a_bydirectcomparison}
    \\
    \bm b=&\beta_0\delta\bm v_n,
    \\
    c=&\frac{\beta^2}{2m}\delta\theta.
    \label{appendix:coef_c_bydirectcomparison}
\end{align}

Since the density and pressure are given by 
\begin{align}
    \tilde n=\frac{g_{3/2}(z)}{\Lambda^3},\\
    \tilde P=\frac{g_{5/2}(z)}{\beta\Lambda^3},
\end{align}
one can expand them to the first order in the fluctuation, given by 
\begin{align}
    \delta\tilde n=\frac{3\tilde n_0}{2}\frac{\delta \theta}{\theta_0}+\gamma_0\theta_0\frac{\delta z}{z_0}, 
    \label{appendix:fluctuationofnoncondensatedensity}
    \\
    \delta\tilde P=\frac{5\tilde P_0}{2}\frac{\delta \theta}{\theta_0}+\tilde n_0\theta_0\frac{\delta z}{z_0}, 
    \label{appendix:fluctuationofpressure}
\end{align}
where we used the equilibrium distribution function for the thermodynamic quantities to evaluate the expansion coefficients.
Using this equation, one can rewrite $\delta\theta$ and $\delta z$ in terms of $\delta \tilde n$ and $\delta \tilde P$.
Using the definition of the fugacity, we can also have the relation 
\begin{align}
    \frac{\delta z}{z_0}=\frac{1}{\theta}\bqty{-(\tilde\mu_0-2gn_0)\frac{\delta\theta}{\theta_0}+\delta\tilde\mu-2g\delta n}.
    \label{appendix:fluctuationoffugacity}
\end{align}
One can solve Eq.~\eqref{appendix:fluctuationoffugacity} with respect to $\delta\tilde\mu$ by making use of Eqs.~\eqref{appendix:fluctuationofnoncondensatedensity} and \eqref{appendix:fluctuationofpressure}.
Substituting the results into Eqs.~\eqref{appendix:coef_a_bydirectcomparison} and~\eqref{appendix:coef_c_bydirectcomparison}, one finds
\begin{align}
    a=&\frac{\beta_0\sigma_{40}}{\tilde n_0}\delta\tilde P+\frac{5\beta_0\sigma_{30}}{2\Gamma_0}\delta\tilde n, 
    \\
    c=&\frac{\beta^2_0}{2m} \left ( 
    \sigma_{30}\frac{\delta\tilde P}{\tilde P_0}-\sigma_{40}\frac{\delta\tilde n}{\tilde n_0}
    \right ),
\end{align}
where we used the nondimensional hydrodynamic coefficients defined in Eqs.~\eqref{sig1},~\eqref{sig2},~\eqref{sig3} and ~\eqref{sig4}.

On the other hand, taking a moment directly on Eq.~\eqref{appendix:leqsolbyexpansion}, we find
\begin{align}
    \expval{\nu^{\rm leq}}=&aW_0+cW_2,\\
    \expval{\bm p\nu^{\rm leq}}=&\frac{1}{3}\bm b W_2,\\
    \expval{p^2\nu^{\rm leq}}=&aW_2+cW_4.
\end{align}
Using the expansion form given in Eq.~\eqref{appendix:sonineandsphericalharmonicsexpansion} and taking the moments, one can also show that $\expval{p^n\nu}=\expval{p^n\nu_0}$ and $\expval{p^{n+1}\cos\theta\nu}=\expval{p^{n+1}\nu_1}/3$.
Thus, imposing the local equilibrium condition $\expval{\nu}=\expval{\nu^{\rm leq}},\expval{\bm p\nu}=\expval{\bm p\nu^{\rm leq}}$ and $\expval{\tilde\varepsilon(p)\nu}=\expval{\tilde\varepsilon(p)\nu^{\rm leq}}$, we find
\begin{align}
    \expval{\nu_0}= & \delta\tilde n,\label{appendix:density_nu0}
    \\
    \expval{p\nu_1}= & 3m\tilde n_0\delta v_n,\label{appendix:vel_p1nu1}
    \\
    \expval{p^2\nu_0}= & 3m\delta\tilde P.\label{appendix:P_p2nu0}
\end{align}

\bibliographystyle{apsrev4-1}
\bibliography{ProofreadMaintext}

\end{document}